\documentclass{aa}  

\usepackage{graphicx}
\usepackage{txfonts}
\usepackage{amsmath}
\usepackage{comment}
\usepackage{color}

\newcommand{\Msun}{{\rm M_{\odot}}}

\newcommand{\Zsun}{{\rm Z_{\odot}}}
\newcommand{\mpc}{\, {\rm Mpc}}
\newcommand{\kpc}{\, {\rm kpc}}
\newcommand{\pc}{\, {\rm pc}}

\newcommand{\kmps}{\, {\rm km \, s^{-1}}}

\begin{document}

   \title{Compact and high excitation molecular clumps in the extended ultraviolet disk of M83}

   \author{Jin Koda\inst{1,2},
   Fran\c{c}oise Combes\inst{3},
   Monica Rubio\inst{4},
   Morten Andersen\inst{5},
   Frank Bigiel\inst{6},
   Armando Gil de Paz\inst{7,8},
   Junais\inst{9},
   Amanda M Lee\inst{10},
   Jennifer Donovan Meyer\inst{11},
   Kana Morokuma-Matsui\inst{12}
   Masafumi Yagi\inst{13},
   Annie Zavagno\inst{14}
          }

   \institute{Observatoire de Paris, LERMA, PSL Univ., CNRS, Sorbonne Univ., Paris, France \and
              Department of Physics and Astronomy, Stony Brook University, Stony Brook, NY 11794-3800, USA \and
              Observatoire de Paris, LERMA, Coll\`ege de France,  PSL Univ., CNRS, Sorbonne Univ., Paris, France \and
              Departamento de Astronomia, Universidad de Chile, Casilla 36-D, Santiago, Chile \and
              European Southern Observatory, Karl-Schwarzschild-Str. 2, 85748 Garching, Germany \and
              Argelander-Institut fu\"r Astronomie, Universita\"t Bonn, Auf dem Hu\"gel 71, D-53121 Bonn, Germany \and
              Departamento de F\'{\i}sica de la Tierra y Astrof\'{\i}sica, Facultad de CC$.$ F\'{\i}sicas, Universidad Complutense de Madrid, 28040, Madrid, Spain \and
              Instituto de F\'{\i}sica de Part\'{\i}culas y del Cosmos IPARCOS, Facultad de CC$.$ F\'{\i}sicas, Universidad Complutense de Madrid, 28040 Madrid, Spain \and
              National Centre for Nuclear Research, Pasteura 7, 02-093 Warsaw, Poland \and
              Department of Astronomy, University of Massachusetts Amherst, 710 North Pleasant Street, Amherst, MA 01003, USA \and
              National Radio Astronomy Observatory (NRAO), 520 Edgemont Road, Charlottesville, VA 22903, USA \and
              Institute of Astronomy, Graduate School of Science, The University of Tokyo, 2-21-1 Osawa, Mitaka, Tokyo 181-0015, Japan \and
              National Astronomical Observatory of Japan, Mitaka, Tokyo, 181-8588, Japan \and
              Aix Marseille Univ., CNRS, CNES, Laboratoire d'Astrophysique de Marseille, Marseille, France
             }
   \date{Received July 15, 2024; accepted September 23, 2024}

    \titlerunning{Compact molecular clumps in the XUV disk of M83}
    \authorrunning{Koda et al.}

  \abstract
   {The extended ultraviolet (XUV) disks of nearby galaxies show ongoing massive-star formation, but their parental molecular clouds remain mostly undetected despite searches in CO(1-0) and CO(2-1).
   The recent detection of 23 clouds in the higher excitation transition CO(3-2) within the XUV disk of M83 thus requires an explanation.} % Context
   {We test the hypothesis introduced to explain the non-detections and recent detection simultaneously:
   The clouds in XUV disks have a clump-envelope structure similar to those in Galactic star-forming clouds, having dense star-forming clumps (or concentrations of multiple clumps) at their centers, which predominantly contribute to the CO(3-2) emission and are surrounded by less dense envelopes, where CO molecules are photo-dissociated due to the low-metallicity environment there.} % Aims
   {We utilized new high-resolution ALMA CO(3-2) observations of a subset (11) of the 23 clouds in the XUV disk of M83.} % Methods
   {We confirm the compactness of the CO(3-2)-emitting dense clumps (or their concentrations), finding clump diameters below the spatial resolution of 6-9~pc. This is similar to the size of the dense gas region in the Orion A molecular cloud, a local star-forming cloud with massive-star formation.} % Results
   {The dense star-forming clumps are common between normal and XUV disks.
   This may also indicate that once the cloud structure is set, the process of star formation is governed by the cloud's internal physics rather than by external triggers.
   This simple model explains the current observations of clouds with ongoing massive-star formation, although it may require some adjustment, for example including the effect of cloud evolution, to describe star formation in molecular clouds more generally.
   } % Conclusions
   \maketitle

\section{Introduction} \label{sec:intro}

The Galaxy Evolution Explorer (GALEX) satellite has enabled the discovery of a surprisingly large number of massive-star formation (SF) sites beyond the optical radii ($R_{25}$) of galactic disks \citep[often $\sim$2-4 $R_{25}$; ][]{Gil-de-Paz:2005aa, Thilker:2005aa, Bigiel:2010yq}.
Numerous bright UV sources (OB stars) are present beyond the edges of the optical disks
and extend into the low-density HI gas disks \citep[see the example of M83 in][]{Bigiel:2010yq, Koda:2012aa, Koda:2022aa, Eibensteiner:2023aa, Rautio:2024aa}.
Such extended ultraviolet (XUV) disks are found in about one-third of local disk galaxies \citep{Thilker:2007dp}.
Thus, this mode of SF is fairly common despite the extreme conditions.
It is important to study the parental sites of this SF activity: the molecular gas clouds.

Efforts have been made to detect CO emission (molecular gas) in XUV disks
but have rarely succeeded (see \citealt{Watson:2017aa} for a review, including unpublished non-detections).
Such studies utilized CO($J$=1-0) or CO($J$=2-1) emission, widely used tracers of the bulk molecular gas
in normal molecular clouds, as these lines are excited easily at the average density ($n_{\rm H_2}\sim 300\,\rm cm^{-3}$) and
temperature ($T_{\rm k}\sim 10\,\rm K$) of the bulk gas.
Until recently, CO was detected in the outer disk environment of only four galaxies 
\citep{Braine:2004aa, Braine:2007aa, Braine:2010fr, Dessauges-Zavadsky:2014fk},
and only one of the four is identified as a single molecular cloud \citep{Braine:2012aa}.
Even a sensitive search in CO(2-1) with the Atacama Large Millimeter/submillimeter Array (ALMA) over a $2\times 4\kpc^2$ region in the XUV disk of M83
resulted in a non-detection \citep{Bicalho:2019aa}.
The rarity of detections in CO(1-0) and CO(2-1) has been at odds with the abundance of UV sources (indicating massive SF) across these XUV disks.

A breakthrough has come recently. 
ALMA revisited the XUV disk of M83, a $1\kpc^2$ region \citep{Koda:2022aa}
within the larger region of the non-detection in CO(2-1) \citep{Bicalho:2019aa}.
It detected 23 molecular clouds in CO(3-2) emission.
While these clouds are spatially unresolved at 1" resolution (20~pc),
the detection led to a hypothesis, presented in \citet{Koda:2022aa}, regarding the parental clouds
of the outer disk massive SF (see Sect. \ref{sec:hypothesis}).
This work aims to test this hypothesis
with higher-resolution CO(3-2) observations,
in conjunction with a consideration of the previous CO(2-1) non-detection.

This hypothesis assumes that the clouds have a warm, dense gas concentration surrounded by an envelope of more diffuse molecular gas (Sect. \ref{sec:hypothesis}).
We refer to this dense gas concentration as a ``clump," though it could be the clump often discussed in resolved Galactic studies with a typical size of 0.3-3~pc \citep{Bergin:2007aa} or a concentration of multiple such clumps of a larger size.

M83 is one of the closest XUV disks at $d=4.5\mpc$ \citep{Thim:2003aa}.
This prototype XUV disk has been a target of optical spectroscopy, having the low metallicity of $\sim1/3\Zsun$ throughout its XUV disk \citep{Gil-de-Paz:2007lj, Bresolin:2009ce}.

\section{Hypothesis and predictions} \label{sec:hypothesis}

\begin{figure*}[h]
    \centering
    \includegraphics[width=1.0\textwidth]{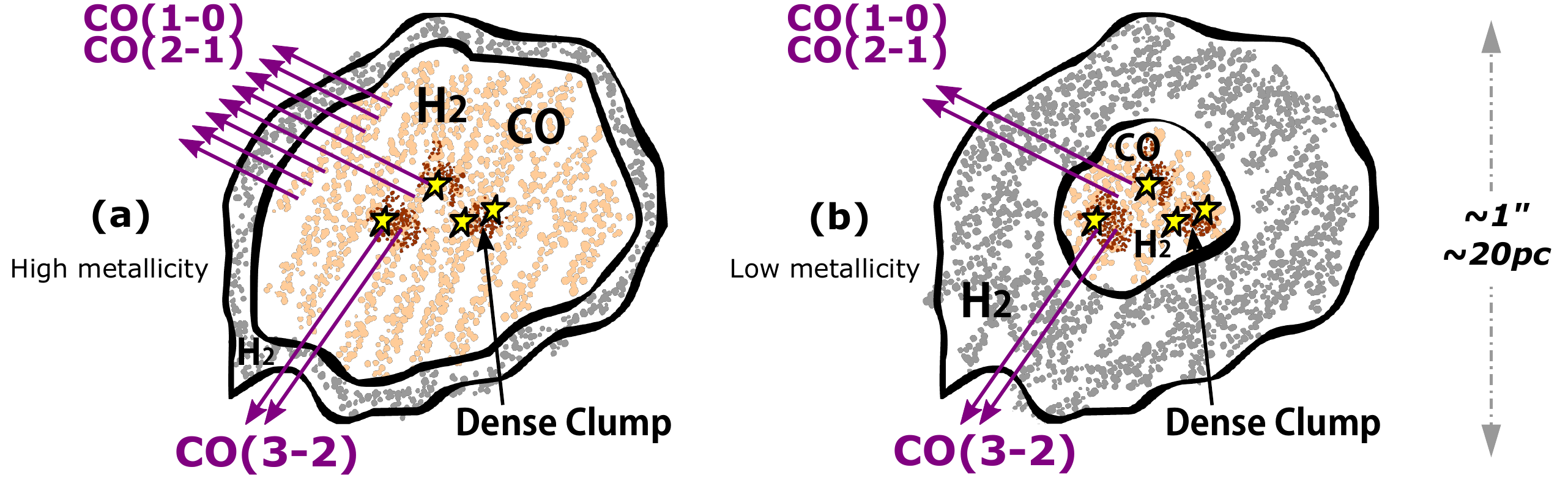}
    \caption{Clump-envelope structure of a molecular cloud in an environment of (a) high metallicity and (b) low metallicity, suggested by \citet{Koda:2022aa}
    based on previous studies \citep[e.g.,][]{Maloney:1988lr, Wolfire:2010aa}.
    In the low-metallicity environment (b), CO molecules are selectively photo-dissociated in the envelope (orange) by the external UV radiation field, producing a thick CO-deficient H$_2$ layer (gray), which substantially reduces the CO(2-1) and CO(1-0) emission.
    The dense star-forming clumps (brown and their immediate surroundings) reside at the hearts of the clouds; the CO molecules are protected there, and the clumps can remain bright in CO(3-2).}
    \label{fig:schematic}
\end{figure*}

Figure \ref{fig:schematic} presents the hypothesis \citep[from ][ with some additional explanations]{Koda:2022aa}.
It assumes that clouds in XUV disks have an internal mass distribution similar to
those of Galactic star-forming clouds \citep[such as the Orion A molecular cloud; e.g.,][]{Ikeda:1999vt, Nakamura:2019wt},
where dense star-forming clumps, or their concentrations,
are embedded in a large envelope of bulk molecular gas (Fig. \ref{fig:schematic}a,b).
In the cloud envelopes, CO molecules can be photo-dissociated, especially 
in low-metallicity environments such as those of XUV disks
\citep[Fig. \ref{fig:schematic}b; see][]{Maloney:1988lr, van-Dishoeck:1988br, Wolfire:2010aa}.
The CO(3-2) excitation has to overcome the high $J$=3 level energy of $E_{J}/k\sim 33\,\rm K$
and high critical density of $\gtrsim 10^3\,\rm cm^{-3}$ (even after photon trapping reduces the critical density).
Thus, the majority of the CO(3-2) emission must be from the dense clumps at the hearts of the clouds,
which can remain CO-rich both in high and low metallicities (Fig. \ref{fig:schematic}a,b).
On the contrary, CO(2-1) can be excited in relatively low $n_{\rm H_2}$ and $T_{\rm k}$
in the cloud envelopes (Fig. \ref{fig:schematic}a).
However, the envelopes are CO-deficient (Fig. \ref{fig:schematic}b) and are dark in CO(2-1)
in the low metallicity of XUV disks.
This can explain the CO(3-2) detection and CO(2-1) non-detection simultaneously.

This hypothesis is testable by two means.
First, the sizes of the dense clumps would be smaller than the typical size of molecular clouds.
For example, the Orion A cloud has a diameter of 20~pc \citep{Nakamura:2019wt}. This size could not be resolved at the 1" (20pc) resolution of the previous CO(3-2) observations in the M83 XUV disk \citep{Koda:2022aa}, but it can be resolved at a higher-resolution in CO(3-2) with the new observations presented in this paper.

Second, the dense clumps should emit CO(2-1) as well, but it is expected to be faint.
If the gas is thermalized at the high $n_{\rm H_2}$ and $T_{\rm k}$ required for the substantial CO(3-2) excitation,
the CO 3-2/2-1 ratio in surface brightness temperature would be close to one, 
and the CO 3-2/2-1 intensity ratio would be $I_{\rm CO(3-2)}/I_{\rm CO(2-1)}=2.25$ (see more detailed calculations in Sect. \ref{sec:excitation}).
This can be tested with deep CO(2-1) observations in conjunction with the measured CO(3-2) fluxes.
Such CO(2-1) observations were approved in ALMA Cy9, but were not executed due largely to
the loss of observing times in small array configurations as a result of the cyberattack in 2022-2023.
In this work, we reevaluate the significance of the upper limit in CO(2-1) flux from the previous non-detection.

\section{Observations and data reduction} \label{sec:obs}

\begin{figure}[h]
    \centering
    \includegraphics[width=0.5\textwidth]{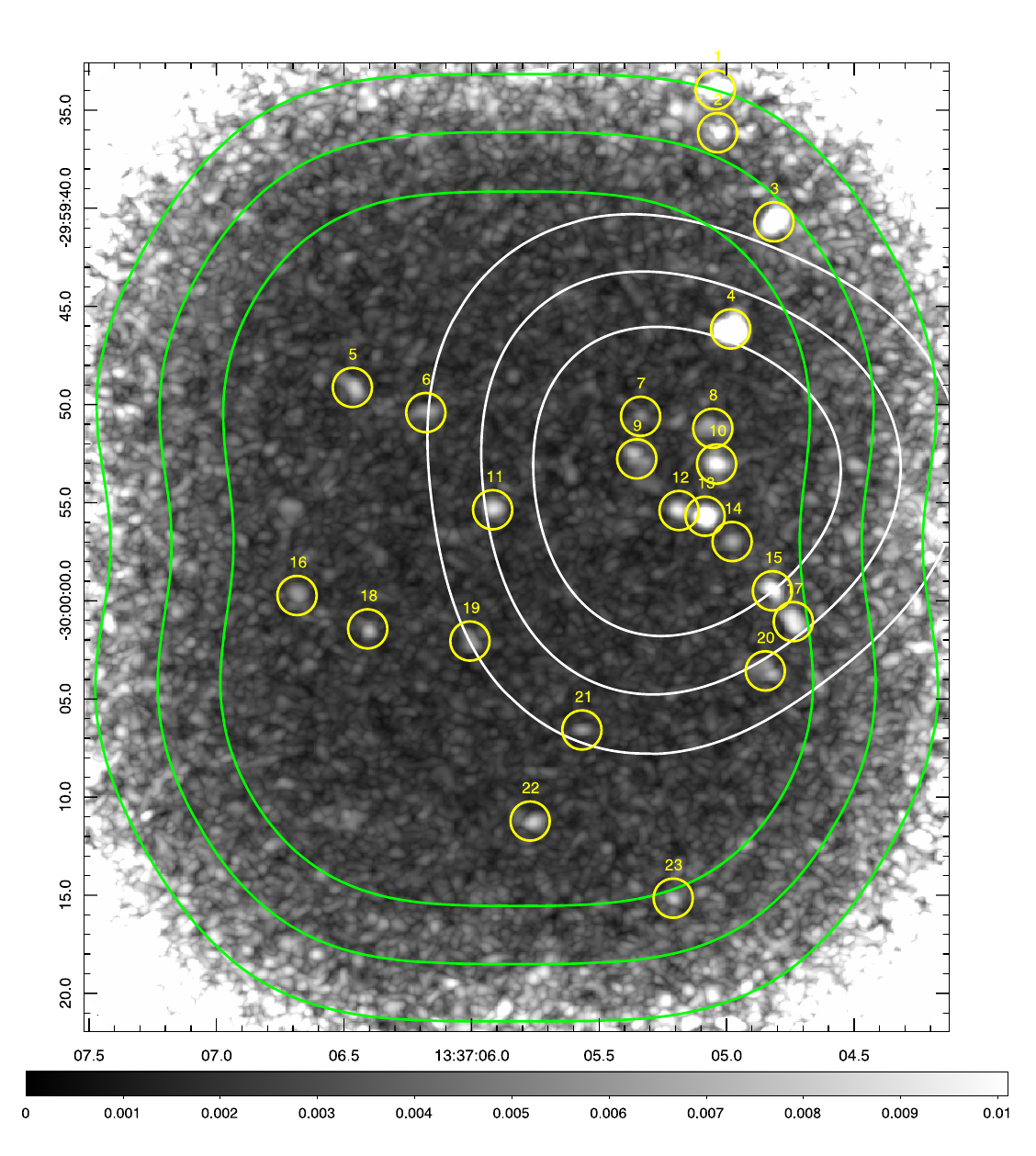}
    \caption{Field of view (FoV) of Cy9 (white contours) with respect to that of Cy7 (green),
    displayed on the CO(3-2) peak intensity map of the Cy7 data imaged with natural weighting (CY7NA).
    The FoV center of Cy9 is at (RA, Dec)$_{\rm J2000}$=(13:37:05.18, -29:59:53.6).
    The sensitivity decreases outward from the centers for both FoVs, and
    the contours show the locations of 70, 50, and 30\% of the peak sensitivity for the Cy7 (green) and 9 data (white).
    The 11 clouds within the 50\% white contour are the focus of this paper.
    The yellow circles have 2" (44~pc) diameters and mark the locations of the 23 molecular clouds in \citet{Koda:2022aa}.
    The colorbar is in units of Jy/beam.
    }
    \label{fig:fov}
\end{figure}

We observed a region with an approximate diameter of 30"
with the ALMA 12-m array in CO(3-2) and dust continuum emission (project \# 2022.1.00359.S in Cycle 9, hereafter Cy9).
It is located at a galactocentric distance of $\sim 1.24 R_{25}$ ($\sim 8.0\arcmin$, $\sim 10.4$~kpc).
The observed region is covered with 3 pointings in CO(3-2).
Figure \ref{fig:fov} shows the region with the sensitivity contours at 70, 50, and 30\%
of the peak sensitivity at the center, which is (RA, Dec)$_{\rm J2000}$=(13:37:05.18, -29:59:53.6).
This region is selected based on previous ALMA observations:
starting from the CO(2-1) observations of a $2 \times 4\kpc^2$ region \citep[][\# 2013.1.00861.S in Cycle 3, Cy3]{Bicalho:2019aa}, and
the $1\kpc^2$ subregion studied in CO(3-2) \citep[the gray scale image in Fig. \ref{fig:fov}; ][\# 2017.1.00065.S in Cycle 7, Cy7]{Koda:2022aa}.
Within the $1\kpc^2$ region, a smaller region is selected for this work.
This region has bright UV and H$\alpha$ peaks and multiple molecular clouds (Fig. \ref{fig:fov}).

We analyzed the Cy3, Cy7, and Cy9 data.
The details of the Cy3 and Cy7 observations are in \citet{Koda:2022aa}.
In the new Cy9 observations, 
we configured one spectral window for the CO(3-2) line emission
with band and channel widths of 937.5~MHz ($812.8\kmps$) and 564.5~kHz ($0.4894\kmps$).
The other three spectral windows were configured to cover different sky frequencies
for the continuum emission with a bandwidth of 1.875~GHz with 128 channels per spectral window.
The central frequency of the continuum emission in the combination of the Cy7 and Cy9 data is 351.3~GHz (hereafter, 351~GHz continuum emission).
The final $uv$-coverage at the CO(3-2) frequency ranges over angular scales of 0.23-12.2$\arcsec$ ($\sim$5.0-266~pc).

The visibility data are calibrated using the data reduction scripts provided by the ALMA Observatory
using the Common Astronomy Software Application \citep[CASA; ][]{McMullin:2007aa, CASA-Team:2022aa}.
The amplitude and phase of the bandpass and gain calibrators are confirmed to be flat over time and frequency after the calibrations.

For imaging, we used the Multichannel Image Reconstruction, Image Analysis, and Display package \citep[MIRIAD;][]{Sault:1995kl, Sault:1996uq}.
The CO(3-2) data from Cy7 and Cy9 were imaged separately and together.
We used the natural (NA) and uniform (UN) weightings.
The parameters for the imaging and final data cubes are listed in Table \ref{tab:data}.
Depending on the dataset and weighting, we refer to each data cube as CY7NA, CY7+9NA, CY7+9UN, and so on (see Table \ref{tab:data}).
The RMS noise per channel is calculated using the dirty cubes before deconvolution in the areas where the primary beam attenuation is practically negligible (PB$>0.95$).
Since the sensitivity and beam shape vary in an asymmetric way across the field of the Cy7 and Cy9 combined data, CY7+9NA and CY7+9UN, they are used only for reference, but not for measurements.

Table \ref{tab:data} includes the coefficient to convert intensity/flux units from Jy/beam to K.
In this paper we mainly use Jy/beam since the sources are not resolved (only marginally resolved at best; see Sect. \ref{sec:sizes}).

\begin{table*}
\caption{Parameters of the reduced data cubes.}             % title of Table
\label{tab:data}
\centering
\begin{tabular}{cccccccccc}
\hline\hline                 % inserts double horizontal lines
(1) & (2) & (3) & (4) & (5) & \multicolumn{3}{c}{(6)} & (7) & (8)\\
\hline
Data & Name &  $\Delta v$ & Cell & Wt. & \multicolumn{3}{c}{Beam Size} & RMS ($1\sigma$) & $C_{\rm K}$ \\
\cline{6-8}
 & & & & & $b_{\rm maj}$, $b_{\rm min}$, PA & \multicolumn{2}{c}{$\sqrt{b_{\rm maj}b_{\rm min}}$} & & \\
\cline{7-8}
 & & $\arcsec$ & $\arcsec$ & & $\arcsec \times \arcsec$, $\deg$ & $\arcsec$ & $\pc$ & mJy/bm & K/[Jy/bm]\\% & mK \\
\hline                        % inserts single horizontal line
Cy3 CO(2-1)   & CY3NA     & 2.54 & 0.15 & NA &  0.75, 0.54, -87.47 & 0.64 & 14  & 10.7 &  56.8  \\% &  607 \\
\hline
Cy7 CO(3-2)   & CY7NA     & 2.54 & 0.05 & NA &  0.96, 0.82, -84.05 & 0.89 & 19  & 1.05 &  13.0    \\% &  13.6 \\
              &           & 0.85 & 0.05 & NA &  0.96, 0.82, -84.05 & 0.89 & 19  & 1.36 &  13.0    \\% &  17.6 \\
\hline
Cy7\&9 CO(3-2)& CY7+9NA   & 2.54 & 0.05 & NA &  0.73, 0.67, -76.31 & 0.70 & 15  & 0.75 &  20.9    \\% &  15.7 \\
\hline
Cy7\&9 CO(3-2)& CY7+9UN   & 2.54 & 0.05 & UN &  0.62, 0.57, -73.89 & 0.59 & 13  & 1.19 &  28.9  \\% &  34.6 \\
\hline
Cy9 CO(3-2)   & CY9NA     & 2.54 & 0.05 & NA &  0.41, 0.38, -49.83 & 0.40 & 8.6 & 0.98 &  65.6   \\% &  64.1 \\
              &           & 0.85 & 0.05 & NA &  0.41, 0.38, -49.83 & 0.40 & 8.6 & 1.44 &  65.6   \\% &  94.2 \\
\hline
Cy9 CO(3-2)   & CY9UN     & 2.54 & 0.05 & UN &  0.30, 0.26, -44.73 & 0.28 & 6.1 & 1.29 & 131.  \\% &  170 \\
              &           & 0.85 & 0.05 & UN &  0.30, 0.26, -44.73 & 0.28 & 6.1 & 1.90 & 131. \\% &  250 \\
\hline
Cy7\&9 Cont.  & CY7+9CONT &      & 0.05 & NA &  0.77, 0.70, -77.56 & 0.73 & 16  & 2.35E-2 &  18.4 \\% &  0.432  \\
\hline                                   %inserts single line
\end{tabular}
\tablefoot{
(1) Data used.
(2) Data name.
(3) Velocity channel width.
(4) Spatial cell/pixel size.
(5) Adopted weighting scheme (NA=Natural, UN=Uniform weighting).
(6) Convolution beam size.
(7) RMS noise. 
The RMS is calculated in region where the primary beam attenuation, with respect to its peak across the field, is $>0.95$.
(8) Conversion coefficient from Jy/beam to K.}
\end{table*}

\section{Cloud and clump identifications and parameters} \label{sec:identification}

We identified clouds and clumps hierarchically, first finding objects in the CY7NA cube
and then searching for their substructures in the CY9NA cube by limiting the search within the volumes of the CY7NA objects,
doing the same with the CY9UN cube within the volumes of the CY9NA objects.
We adopted this approach because the CY7NA data have the highest sensitivity, albeit the lowest spatial resolution (see Table \ref{tab:data}).
Once the volumes of the objects were restricted by the CY7NA analysis, we relaxed search thresholds in the CY9NA and CY9UN data that have higher resolutions but lower sensitivities (see the details below).
To remove the effects of sensitivity variations across the observed fields, we generated and used signal-to-noise ratio (S/N) cubes for the identifications.
Unless stated otherwise, we adopted the cubes with a $2.54\kmps$ channel width.

We refer to the objects identified in the CY7NA cube as clouds and those in the CY9NA and CY9UN cubes as clumps.
These names reflect only the spatial resolutions of the data used for their identifications.
The CY7NA cube has a resolution of 0.89" (19~pc), approximately the diameter of a molecular cloud, and the CY9NA and CY9UN cubes have 0.40" (8.6~pc) and 0.28" (6.1~pc), respectively, tracing sub-cloud scales.

\subsection{Clouds} \label{sec:cloudidentification}
Twenty-three molecular clouds were previously identified by \citet{Koda:2022aa} in the same Cy7 data with an earlier data reduction.
For the sake of consistency, we repeated the same identification procedure on the new CY7NA data ($2.54\kmps$ channel width).
In the S/N cube, we first find pixels with $>5\sigma$ significance and extend their volumes down to pixels with $3\sigma$ per channel.
In some cases, the envelopes of two clouds overlap, which are split with the watershed algorithm \citep[\textsc{Clumpfind};][]{Williams:1994cq}.
We find the same set of 23 molecular clouds as in \citet{Koda:2022aa}.

This cloud identification defines a volume of each cloud in the three-dimensional data cubes, which is used as a mask for the clump identification in the next subsection.

Hereafter, we discuss only the clouds within the Cy9 field of view (to the 50\% level of the primary beam), which includes eleven clouds (Clouds 4, 7, 8, 9, 10, 11, 12, 13, 14, 15, and 17; see Fig. \ref{fig:fov}).
Figures \ref{fig:pstamp0} and \ref{fig:pstamp8} show their cutout images in integrated intensity and peak intensity, respectively.
The clouds are arranged from left to right, with their ID\#s in the top-left corners.
They are also shown from several different imaging runs in CY7NA, CY7+9NA, CY7+9UN, CY9NA, and CY9UN.
These 11 clouds are listed in Table \ref{tab:measured} (the full list of the 23 clouds is in Table \ref{tab:allclouds}).
Cloud ID is labeled in the format ``XX."

\begin{figure*}[h]
    \centering
    \includegraphics[width=1.0\textwidth]{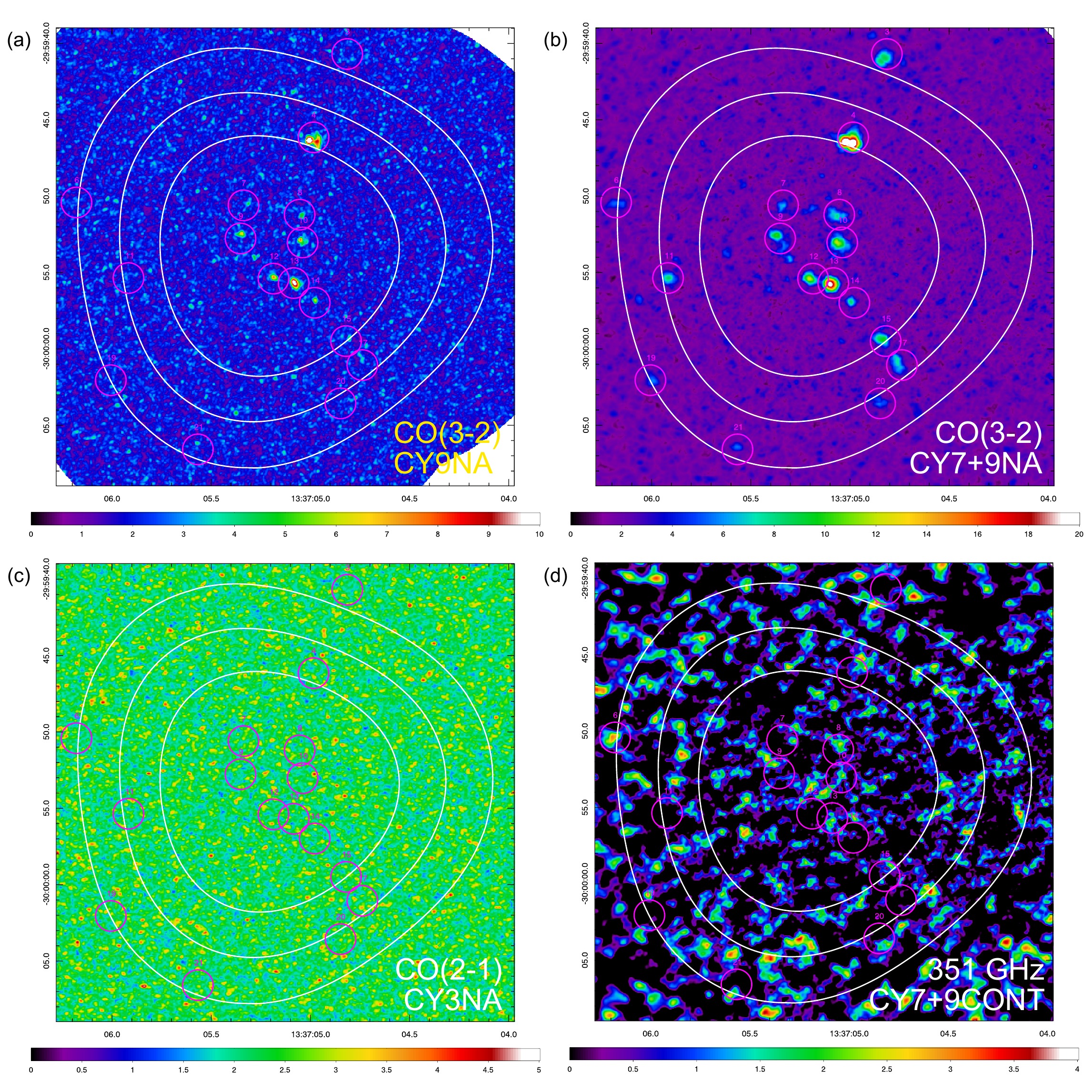}
    \caption{S/N images in peak intensity: (a) CY9NA - CO(3-2), (b) CY7+9NA - CO(3-2), (c) CY3NA - CO(2-1), and (d) CY7+9CONT - 351~GHz continuum.
    The white contours show the locations of 70, 50, and 30\% the peak sensitivity for the Cy9 data.
    The magenta circles have diameters of 2" (44~pc) and mark the locations of the clouds; the cloud IDs are given above each circle.
    }
    \label{fig:eximages}
\end{figure*}

None of the eleven clouds is above the $3\sigma$ detection limit in the CO(2-1) line and 351~GHz continuum emission.
Figure \ref{fig:eximages} shows the S/N images calculated from the peak intensity maps: (a)(b) CO(3-2), (c) CO(2-1), and (d) 351~GHz continuum.
The CO(2-1) emission is not convincingly detected even in a stack analysis of the 11 cutout images (see Appendix \ref{sec:stack}).

\subsection{Clumps (Level A and B)} \label{sec:clumpidentification}

We first searched the CY9NA cube, finding clumps within the cloud volumes identified in CY7NA (in the CY7NA mask),
and then searched the CY9UN cube to find clumps within the volumes of the CY9NA clumps.
To differentiate them, the first group of CY9NA clumps is called "Level A clumps," and the second group, from CY9UN, is "Level B clumps."
Level B clumps reside within the volumes of Level A clumps and are their substructures.
We adopted the same identification procedure as for the clouds (Sect. \ref{sec:cloudidentification})
but limited the search volumes.

For Level A clumps, we started from the CY9NA cube and applied the CY7NA mask to limit the search range within the volume of the CY7NA clouds.
Since this is the volume where a cloud exists at a higher significance, it is less likely to pick up noise even with a lenient detection threshold.
Hence, we started from $>4\sigma$ peaks and extended them to the $3\sigma$ envelopes.
In the CY9NA cube, we detect one or two clumps within all the clouds except Clouds 11 and 17.
These two clouds are near the edge of the Cy9 field of view (Fig. \ref{fig:fov}),
where the sensitivities are compromised due to the primary beam attenuation.
In Table \ref{tab:measured}, the eleven CY9NA clumps are labeled in the format ``XX-YY" with ``XX" as the parental cloud ID and ``YY" as the clump ID within the cloud.

For Level B clumps, we did the same as for the CY9UN cube, using the same detection thresholds, but within the volumes of the CY9NA clumps as masks.
The CY9UN cube has the lowest sensitivity among the CO(3-2) data cubes (Table \ref{tab:data}) and can identify only the brightest clumps, and yet at the highest spatial resolution.
We find nine CY9UN clumps.
They are listed in Table \ref{tab:measured} with the IDs labeled in the format ``XX-YY-ZZ."

Clouds 4 and 10 are resolved to two clumps 4-1 and 4-2, and 10-1 and 10-2, respectively.
Clump 4-2 is resolved further into three clumps, 4-2-1, 4-2-2, and 4-2-3.

\subsection{Parameters} \label{sec:parameters}

Table \ref{tab:measured} (and \ref{tab:allclouds}) lists the measured parameters of the clouds and clumps, including their ID; positions in RA, Dec, and velocity; full width half maximum (FWHM) sizes in RA (=x), Dec (=y), and recession velocity (=v) directions; peak brightness temperature $T_{\rm p}$, peak intensity $I_{\rm p}$ and its uncertainty $\Delta I_{\rm p}$; the integrated flux $F$ and its uncertainty $\Delta F$; the velocity dispersion $\sigma_{\rm v}$ and its uncertainty $\Delta \sigma_{\rm v}$; and the primary beam attenuation (the fraction with respect to the peak sensitivity).
We used the CY7NA, CY9NA, and CY9UN cubes and masks for the measurements of the CY7NA clouds and the CY9NA and CY9UN clumps, respectively.
We adopted the cubes of the $2.54\kmps$ channel width for these measurements except for ${\rm FWHM}_{\rm v}$ (and $\sigma_{\rm v}$), for which we used $0.85\kmps$ cubes.

As explained, cloud and clump ID are labeled in a hierarchical manner.
The substructures inherit the labels of their parental structures: ``XX" for the CY7NA clouds, ``XX-YY" for the CY9NA (Level A) clumps, and ``XX-YY-ZZ" for the CY9UN (Level B) clumps.

The coordinates are the intensity-weighted centroid positions.

The $N_{\rm pix}$ is the number of the pixels identified in a cloud or clump in the RA-Dec-Vel volume (i.e., x-y-v volume).
$N_{\rm xy}$ is the number of the pixels in the RA-Dec projection (x-y plane).
Some clumps show $N_{\rm pix}=N_{\rm xy}$, meaning that they are detected in one velocity channel.
For reference, the areas of the beams (Table \ref{tab:data}) correspond to $N_{\rm xy}$=356.8 (CY7NA), 70.6 (CY9NA), and 35.4 (CY9UN), respectively.

The FWHMs are calculated from the dispersions $\sigma_{\rm i}^{\rm obs}$ (where $i$=x, y, v) within the cloud volume in the cube, assuming Gaussian profiles in space and velocity, as
\begin{equation}
    {\rm FWHM}_i = \sqrt{8 \ln 2} \, \sigma_{\rm i}^{\rm obs}.
\end{equation}
Table \ref{tab:measured} lists ${\rm FWHM}_i$s instead of $\sigma_{\rm i}^{\rm obs}$, because they are directly comparable to the beam sizes.
We used the $0.85\kmps$ cubes for ${\rm FWHM}_{\rm v}$; we re-gridded the masks from the $2.54\kmps$ cubes and applied them to the $0.85\kmps$ cubes before taking the measurements.

We note that the beam sizes are different for the CY7NA clouds, CY9NA clumps, and CY9UN clumps (Table \ref{tab:data}).
FWHMs in Tables \ref{tab:measured} and \ref{tab:allclouds} are the measured values before dilution corrections for the beam size and velocity resolution.

The $I_{\rm p}$ and $F$ are the values corrected for primary beam attenuation.
$I_{\rm p}$ is also written in the form of the brightness temperature, $T_{\rm p}$, in kelvins, converted with the Rayleigh-Jeans equation.
$\Delta I_{\rm p}$ is from the RMS noise in Table \ref{tab:data}.
$\Delta F$ was calculated with the standard noise propagation starting from $\Delta I_{\rm p}$ by taking into account the number of beams in the volume of objects.

The S/N ($I_{\rm p}/\Delta I_{\rm p}$) can be calculated from Table \ref{tab:measured}.
Level A (CY9NA) clumps are detected at peak S/N ratios of 4-15, mostly at S/N$>5$ with the exceptions of Clumps 10-2, 15-1, and 7-1 (S/N=4-5).
Level B (CY9UN) clumps are detected less significantly, mostly in the range S/N=4-6 except 4-1-1 (S/N=9.9).
Figures \ref{fig:pstamp0} and \ref{fig:pstamp8} show their integrated intensity and peak intensity maps.

From these measured parameters, we calculated the radius, $R,$ and velocity dispersion, $\sigma_{\rm v}$, of the clouds and clumps.
The observed $\sigma_{\rm i}^{\rm obs}$ and ${\rm FWHM}_i$ contain the beam size or velocity channel width.
To remove their effects, we calculated the second moments in radius and velocity as
\begin{eqnarray}
    \sigma_{\rm r} &=& \sqrt{\sigma_{\rm x}^{\rm obs} \sigma_{\rm y}^{\rm obs} - \frac{b_{\rm maj}b_{\rm min}}{8 \ln 2}}, \label{eq:sigr} \\
    \sigma_{\rm v} &=& \sqrt{ (\sigma_{\rm v}^{\rm obs})^2 - \sigma_{\rm ch}^2}, \label{eq:sigv}
\end{eqnarray}
and
\begin{equation}
    R=1.92\sigma_{\rm r}. \label{eq:radius}
\end{equation}
The $\sigma_{\rm ch}^2$ is the second moment of the channel window function (channel shape),
and $\sigma_{\rm ch}=w/\sqrt{12}$ for a top-hat function with a full width of $w$.
We measured the $\sigma_{\rm v}$ and ${\rm FWHM}_{\rm v}$ with the cubes of $w=0.85\kmps$.
For $R$, we used Eq. (\ref{eq:radius}), the definition by \citet{Solomon:1987pr}, which is supposed to represent the whole radial extent of the object.

Equation (\ref{eq:sigr}) requires the measured FWHM$_x$ and FWHM$_y$ to be greater than $b_{\rm maj}$ and $b_{\rm min}$ (beam size) and the  $\sigma_{\rm r}$ and $R$ to be positive.
This condition is met only for Cloud 4 and Clump 13-1, and even their $R$ (=0.35" and 0.20", respectively) are less than half the beam sizes.
Therefore, we consider that none of the clouds nor the clumps are spatially resolved in any of the measurements.
We followed \citet{Koda:2022aa} to set the upper limit of their size and
considered $R$ to be measurable only when the measured size is at least 20\% larger than the beam size.
The minimum measurable radius is then $R_{\rm min}=0.54\sqrt{b_{\rm maj}b_{\rm min}}$.
This is 0.50" (10.9~pc), 0.22" (4.8~pc), and 0.16" (3.5~pc) for the clouds (CY7NA), Level A clumps (CY9NA), and Level B clumps (CY9UN), respectively.
Obviously, these minimum radii correspond to the diameters of 1.00" (21.8~pc), 0.44" (9.6~pc), and 0.32" (7.0~pc), respectively.

The $R$ (upper limit) and $\sigma_{\rm v}$ are listed in Table \ref{tab:measured}.

\section{Small sizes of CO(3-2) clumps} \label{sec:sizes}

Our hypothesis is that the CO(3-2) emitting regions are dense clumps embedded in the CO-deficient cloud envelopes (Fig. \ref{fig:schematic} and Sect. \ref{sec:hypothesis}).
If this is correct, the CO(3-2) emitting regions should be compact compared to typical cloud sizes.
In what follows, we show that the CO(3-2) emitting regions are indeed compact and are not spatially resolved (or at most, are only marginally resolved) in the new data, based on a consideration of the measured sizes, peak intensities, and total fluxes of the clouds and clumps.

\subsection{Size measurements}

Figures \ref{fig:pstamp0} and \ref{fig:pstamp8} show the clouds and clumps, from the top to bottom panels, at the lowest (0.89", 19~pc for CY7NA) to highest resolutions (0.40", 8.7~pc for CY9NA and 0.28", 6.1~pc for CY9UN).
The yellow circles have a diameter of 2" (44~pc),
which is close to a typical diameter of giant molecular clouds, 40~pc, in the Milky Way \citep{Scoville:1987vo}.
We recall that the clouds are identified in the CY7NA data, and Level A and B clumps are in the CY9NA and CY9UN data, respectively.

Clouds 4 and 10 are formally resolved into smaller clumps by the identification criteria in Sect. \ref{sec:clumpidentification}.
None of these clumps are resolved at the higher resolutions (Table \ref{tab:measured}).
By eye, clouds 13 and 17 could also be split into smaller clumps, but are not formally resolved with the identification criteria.
Their apparent clumps appear to be as small as those in cloud 4 (Figs. \ref{fig:pstamp0} and \ref{fig:pstamp8}).
The other clouds have only a single, unresolved clump (Table \ref{tab:measured}).

Quantitatively, in Table \ref{tab:measured}, the  FWHM$_x$ and  FWHM$_y$ of the clouds are similar to the beam size of the CY7NA data (0.89", 19~pc).
This linear spatial scale is close to the sizes of small molecular clouds in the Milky Way disk.
 FWHM$_x$ and  FWHM$_y$ of Level A and B clumps are also similar to the beam sizes of CY9NA and CY9UN (0.40", 9~pc and 0.28", 6~pc, respectively).
Hence, they are not spatially resolved even at the higher resolutions.

From this, we conclude that the CO(3-2) emitting clumps in each cloud are not spatially resolved at the beam sizes of CY9NA and CY9UN.
We note that our size measurements are not very accurate because our clump detection thresholds are not high: S/N$>4$ and $>3$ at the peak and envelope.
Within the uncertainties, the CO(3-2) sources appear to be as compact as the beam sizes of $\sim$6-9~pc or possibly smaller.

\subsection{Implications of the peak intensities and fluxes}

Table \ref{tab:ratio} compares the peak intensities (top part) and total fluxes (bottom) between the parental clouds and their clumps (both Level A and B clumps).
Each row gives measurements within each cloud.
The peak intensity of the clumps, $\max (I_{\rm p})$, is that of the brightest clump within the cloud.
The total flux of the clumps, $\sum F$, is the sum of the fluxes of all the identified clumps in the cloud.
Table \ref{tab:ratio} also lists their ratios (i.e., the fractions of emission in the clumps over that in the cloud).

\begin{figure}[h]
    \centering
    \includegraphics[width=0.48\textwidth]{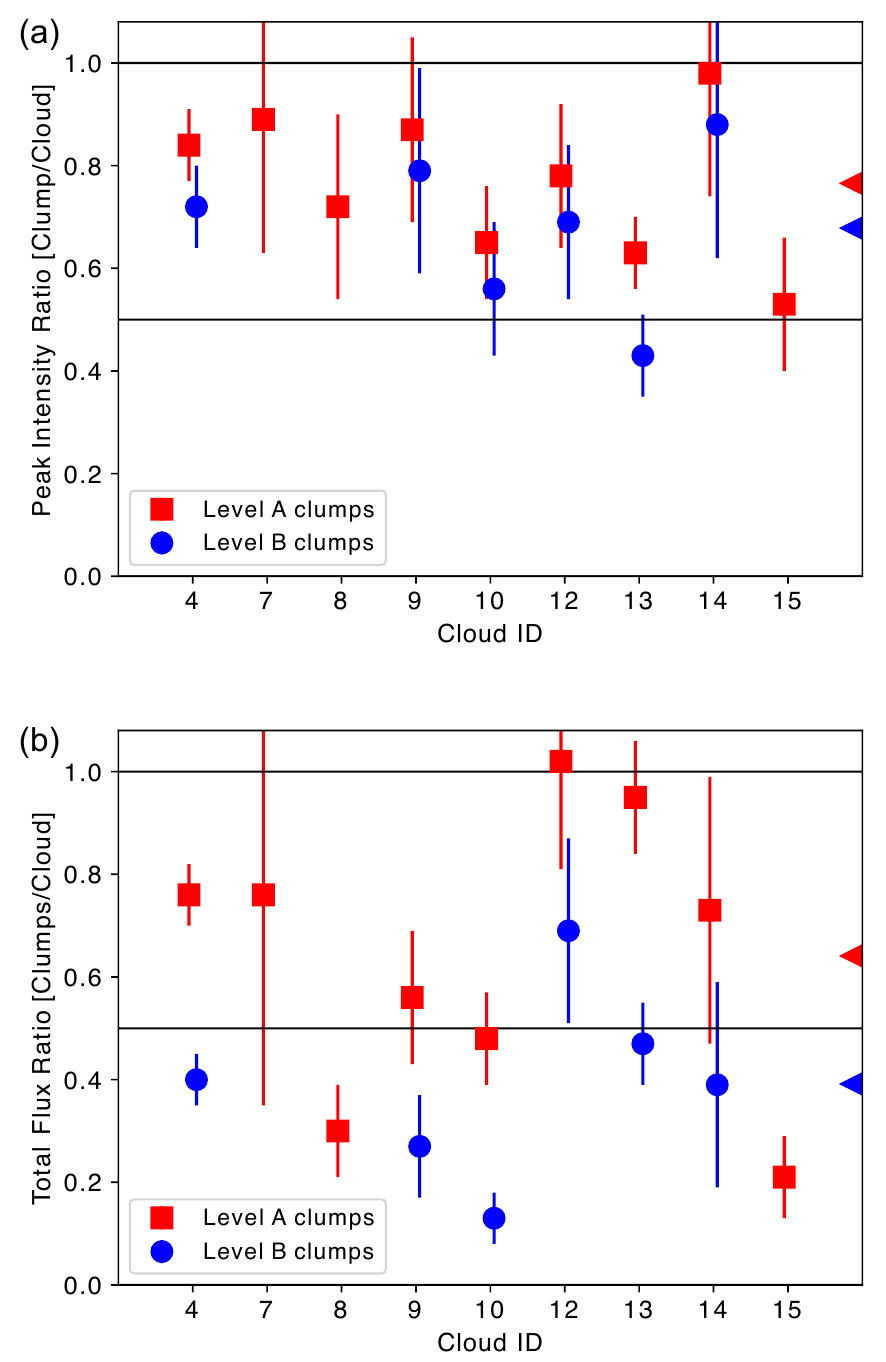}
    \caption{Clumps over cloud ratios for Level A (red square) and B clumps (blue circle) measured in CO(3-2).
    (a) Peak intensity ratio. The peak intensity of the brightest clump, $\max (\mathcal{I}_p)$ in mJy/beam(1), is compared with the peak intensity of its parental cloud, $I_{\rm p}$, in mJy/beam(2).
    Note that the two beams (1 and 2) have different sizes.
    (b) Total flux ratio.
    On the right edge of each panel, the triangles pointing left show the averages of the Level A (red) and B (blue) clumps.}
    \label{fig:ratios}
\end{figure}

Figure \ref{fig:ratios}a shows the peak intensity ratio between the brightest clump in a cloud and the parental cloud (columns 4 and 6 in the top part of Table \ref{tab:ratio}).
The red squares are for Level A clumps, and the blue circles are for Level B clumps.
In all cases, this ratio is as high as $\sim$50-100\% with the averages of 77\% for Level A and 68\% for Level B.
The peak intensity $I_{\rm p}$ represents a flux within a beam area at the peak position within an object; and the beam sizes are different between the cloud (0.89") and clump measurements (0.40" or 0.28").
The high ratios indicate that approximately the same amount of flux is enclosed within the large and small beams, and hence, the CO(3-2) sources are not much spatially resolved even with the small beams.
This supports the hypothesis that the CO(3-2) emission comes from a small region within a cloud.
The resolved clouds (especially \# 10 and 13) have relatively low fractions because the cloud flux is split into the fluxes of the multiple unresolved clumps, and this measurements take only one of them  (the brightest) into account.

Figure \ref{fig:ratios}b shows the fraction of the total flux of all clumps over that in their parental cloud (columns 4 and 6 in the bottom part of Table \ref{tab:ratio}).
An overall trend of this fraction appears different between Level A and B clumps.
For Level A (red squares), the fraction is high and is mostly $\sim$50-100\%, with exceptions of clouds 8 and 15 where it is 30 and 21\%, respectively.
The areal ratio of the CY7NA and CY9NA beams is $(0.40\arcsec/0.89\arcsec)^2\sim 0.2$, and hence, the fraction would be 20\% if the emission is uniformly distributed within the CY7NA's $0.89\arcsec$ beam.
The measured fractions of $\gg 20\%$ suggest that the CO(3-2)-emitting regions are concentrated mostly in the CY9NA's 0.40" beam.
For Level B (blue circles), the fraction decreases to $\sim$10-50\% except for cloud 12.
The decreasing trend from the low to high resolutions \textit{may} indicate that the clumps are starting to be resolved between the resolutions of CY9NA (0.40", 8.7~pc) and CY9UN (0.28", 6.1~pc).
In fact, the fraction of Level B over Level A in total flux is $\sim$30-70\% (column 7 in Table \ref{tab:ratio} bottom).
This is similar to the areal ratio of the CY9NA and CY9UN beams, $(0.28\arcsec/0.40\arcsec)^2\sim 0.5$.
Therefore, the CO(3-2)-emitting clumps may be being resolved at the resolutions of $\sim$0.28-0.40" (6-9~pc).

We should note a couple of caveats.
The above discussion is to capture the average characteristics of the observed clouds and clumps,
and there are clearly some variations among them.
For example, the small total flux fractions of Level A clumps over clouds 8 and 15 may indicate that there are some CO(3-2) emission outside the unresolved clumps.
We also caution that interferometer imaging without short spacing data (or with the uniform weighting) often reduces the apparent flux.
This is because the emission contained in those data is missed, and also because they carry the flux information and their weight reduction causes systematic uncertainties in the imaging process.
However, the overall trends discussed above suggest that, on average,
the CO(3-2)-emitting objects are smaller than the typical sizes of Galactic molecular clouds.

We draw the same conclusion from the peak surface brightness temperature $T_{\rm p}$.
Table \ref{tab:measured} shows that $T_{\rm p}$ generally increases from the low-resolution (CY7NA) to high-resolution data (CY9UN), indicating that the objects are unresolved, but are approaching to be resolved.
The CO emission easily becomes optically thick on theoretical and empirical bases \citep{Goldreich:1974ab, Solomon:1979px},
and thus, we assumed that it is optically thick.
Using the beam filling factor $f$ (i.e., the areal fraction of an object over beam),
$T_{\rm p}$ is related to the excitation temperature $T_{32}$ as $T_{\rm p}= [ f \Gamma(\nu_{32}, T_{32}) ] T_{32}$ (see Eq. \ref{eq:Gamma} for $\Gamma$).
$\Gamma$ is an increasing function of $T_{32}$ and can be evaluated numerically.
This gives $T_{\rm p}= (3.9 f$ - $22.5f)$~K for $T_{32}=$10-30~K with $f$ as an unknown factor.
Compared to $T_{\rm p}$ in Table \ref{tab:measured} (the brightest is Clump 4-1-1 with $T_{\rm p}=2.49$~K), the filling factor is $f=0.64$ or smaller, again indicating unresolved or marginally resolved objects.
The CO emission rarely becomes optically thin, except under some very special conditions (e.g., line wings and extremely low metallicity); thus, we did not consider it further but briefly note that an optically thin case would reduce the apparent $T_{\rm p}$ and, hence, make the above $f$ a lower limit.

\begin{table*}
\caption{Comparison of clouds and Level A and B clumps in terms of peak intensity and total flux.}             % title of Table
\label{tab:ratio}
\centering
\begin{tabular}{ccccccc}
\hline\hline                 % inserts double horizontal lines
(1) & (2) & (3) & (4) & (5) & (6)  & (7)\\
Cloud ID & Cloud & Level A & (3)/(2) & Level B & (5)/(2) & (5)/(3)\\
\hline
\hline 
    &            $I_{\rm p}$  & $\max_{\rm A} (I_{\rm p})$ & $\max_{\rm A} (I_{\rm p})/I_{\rm p}$ & $\max_{\rm B} (I_{\rm p})$ & $\max_{\rm B} (I_{\rm p})/I_{\rm p}$ & $\max_{\rm B} (I_{\rm p})/\max_{\rm A} (I_{\rm p})$ \\
    &            mJy/bm &          (mJy/bm) &                   &            mJy/bm &                   &                    \\
\hline
  4 &  26.47 $\pm$ 1.25 &  22.21 $\pm$ 1.47 &   0.84 $\pm$ 0.07 &  19.04 $\pm$ 1.92 &   0.72 $\pm$ 0.08 &   0.86 $\pm$ 0.10  \\
  7 &   5.53 $\pm$ 1.08 &   4.92 $\pm$ 1.06 &   0.89 $\pm$ 0.26 &                   &                   &                    \\
  8 &   7.26 $\pm$ 1.17 &   5.21 $\pm$ 1.04 &   0.72 $\pm$ 0.18 &                   &                   &                    \\
  9 &   7.80 $\pm$ 1.08 &   6.79 $\pm$ 1.02 &   0.87 $\pm$ 0.18 &   6.16 $\pm$ 1.33 &   0.79 $\pm$ 0.20 &   0.91 $\pm$ 0.23  \\
 10 &  11.64 $\pm$ 1.17 &   7.56 $\pm$ 1.00 &   0.65 $\pm$ 0.11 &   6.49 $\pm$ 1.31 &   0.56 $\pm$ 0.13 &   0.86 $\pm$ 0.19  \\
 12 &   9.74 $\pm$ 1.11 &   7.59 $\pm$ 1.00 &   0.78 $\pm$ 0.14 &   6.70 $\pm$ 1.30 &   0.69 $\pm$ 0.15 &   0.88 $\pm$ 0.20  \\
 13 &  17.57 $\pm$ 1.16 &  11.15 $\pm$ 1.01 &   0.63 $\pm$ 0.07 &   7.63 $\pm$ 1.32 &   0.43 $\pm$ 0.08 &   0.68 $\pm$ 0.13  \\
 14 &   6.81 $\pm$ 1.22 &   6.64 $\pm$ 1.09 &   0.98 $\pm$ 0.24 &   5.99 $\pm$ 1.42 &   0.88 $\pm$ 0.26 &   0.90 $\pm$ 0.27  \\
 15 &  12.09 $\pm$ 1.34 &   6.38 $\pm$ 1.42 &   0.53 $\pm$ 0.13 &                   &                   &                    \\
\hline
    &               $F$ &  $\sum_{\rm A} F$ &                   &  $\sum_{\rm B} F$ &                   &                    \\
    &   $\rm mJy \kmps$ &   $\rm mJy \kmps$ &                   &   $\rm mJy \kmps$ &                   &                    \\
\hline
  4 & 258.20 $\pm$11.20 & 196.00 $\pm$12.60 &   0.76 $\pm$ 0.06 & 104.50 $\pm$11.10 &   0.40 $\pm$ 0.05 &   0.53 $\pm$ 0.06  \\
  7 &   5.80 $\pm$ 2.00 &   4.40 $\pm$ 1.80 &   0.76 $\pm$ 0.41 &                   &                   &                    \\
  8 &  30.50 $\pm$ 4.70 &   9.30 $\pm$ 2.50 &   0.30 $\pm$ 0.09 &                   &                   &                    \\
  9 &  30.20 $\pm$ 4.40 &  17.00 $\pm$ 3.20 &   0.56 $\pm$ 0.13 &   8.20 $\pm$ 2.80 &   0.27 $\pm$ 0.10 &   0.48 $\pm$ 0.18  \\
 10 &  49.60 $\pm$ 5.70 &  23.60 $\pm$ 3.80 &   0.48 $\pm$ 0.09 &   6.30 $\pm$ 2.30 &   0.13 $\pm$ 0.05 &   0.27 $\pm$ 0.10  \\
 12 &  27.40 $\pm$ 4.00 &  28.00 $\pm$ 3.90 &   1.02 $\pm$ 0.21 &  19.00 $\pm$ 4.00 &   0.69 $\pm$ 0.18 &   0.68 $\pm$ 0.17  \\
 13 &  78.60 $\pm$ 6.40 &  74.50 $\pm$ 6.10 &   0.95 $\pm$ 0.11 &  36.90 $\pm$ 5.70 &   0.47 $\pm$ 0.08 &   0.50 $\pm$ 0.09  \\
 14 &  13.30 $\pm$ 3.30 &   9.70 $\pm$ 2.50 &   0.73 $\pm$ 0.26 &   5.20 $\pm$ 2.30 &   0.39 $\pm$ 0.20 &   0.54 $\pm$ 0.27  \\
 15 &  35.80 $\pm$ 5.20 &   7.40 $\pm$ 2.70 &   0.21 $\pm$ 0.08 &                   &                   &                    \\
\hline                                   %inserts single line
 \end{tabular}
 \tablefoot{
 Peak and integrated intensities for clouds and clumps.
 (1) Cloud ID.
 (2) Peak intensity $I_{\rm p}$ or total flux $F$, the same as those in Table \ref{tab:measured}.
 (3)(5) Maximum $I_{\rm p}$ among clumps or total flux $F$ of all clumps for Level A and B clumps.
 (4)(6)(7) Ratios of Level A/Cloud, Level B/Cloud, and Level B/Level A.
 The beam sizes for Cloud ($\sim 0.89\arcsec$), Level A clumps ($\sim 0.40\arcsec$), and Level B clumps  ($\sim 0.28\arcsec$) are different.
}
\end{table*}

\section{Non-detection in CO(2-1)} \label{sec:excitation}

The combination of the detection in CO(3-2) and non-detection in CO(2-1) also constrains the cloud structure and physical condition.
The new data reduction of the CO(2-1) also supports this.

\citet{Koda:2022aa} assumed, as in Fig. \ref{fig:schematic}b, that
[1] the cloud envelope is CO-deficient due to the low metallicity and does not emit much CO emission,
while the dense clump(s) in the cloud is still CO-rich and emit CO(3-2) as well as CO(1-0) and CO(2-1),
and
[2] the CO-rich part of the cloud (i.e., the dense clumps) is close to being thermalized, and hence, the CO 3-2/2-1 ratio in surface brightness temperature is relatively high and is approximately unity.
Under these assumptions, the CO(2-1) flux can be predicted from the observed CO(3-2) flux.
\citet{Koda:2022aa} concluded that the predicted CO(2-1) flux of the brightest cloud
is just below the detection limit of the previous CO(2-1) observations \citep{Bicalho:2019aa}.
We note that in this section, we discuss ``flux" instead of ``surface brightness temperature" because the emission sources are not resolved.

The assumptions that \citet{Koda:2022aa} adopted predict the minimum of the CO(2-1) flux.
In other words, if any of the assumptions deviate significantly from the reality, the predicted CO(2-1) flux would be higher
and should have been detected in the previous CO(2-1) study.

The CO(2-1) flux of a CO(3-2)-emitting dense clump, without the cloud envelope, can be estimated from its CO(3-2) flux.
We assumed that the beam filling factors of the two transitions are equal.
Using the flux $F_i$, excitation temperature $T_i$, and frequency $\nu_i$ of the CO(2-1) and CO(3-2) emission ($i=$21 or 32, respectively), the CO 3-2/2-1 flux ratio is
\begin{equation}
    \frac{F_{\rm 21}}{F_{\rm 32}} 
    = \left( \frac{\nu_{\rm 21}}{\nu_{\rm 32}} \right)^2 \left( \frac{T_{\rm 21}}{T_{\rm 32}} \right) \left( \frac{\Gamma(\nu_{\rm 21}, T_{\rm 21})}{\Gamma(\nu_{\rm 32}, T_{\rm 32})} \right) 
    = \frac{4}{9} \left( \frac{T_{\rm 21}}{T_{\rm 32}} \right) \alpha(T_{\rm 21}, T_{\rm 32}), \label{eq:f21}
\end{equation}
where
\begin{equation}
    \Gamma(\nu, T) = \frac{h\nu/kT}{\exp{(h\nu/kT)}-1}, \label{eq:Gamma}
\end{equation}
and we defined
\begin{equation}
    \alpha(T_{\rm 21}, T_{\rm 32}) = \frac{\Gamma(\nu_{\rm 21}, T_{\rm 21})}{\Gamma(\nu_{\rm 32}, T_{\rm 32})}.
\end{equation}
The ``$T_{\rm 21}/T_{\rm 32}$" term in Eq. (\ref{eq:f21}) takes the minimum under the thermalized condition ($T_{\rm 21}=T_{\rm 32}$) with respect to non-thermalized conditions ($T_{\rm 21}>T_{\rm 32}$).
The ``$\alpha$" term can be evaluated numerically and is always $\alpha(T_{\rm 21}>T_{\rm 32})>\alpha(T_{\rm 21}=T_{\rm 32})$, taking the minimum in the thermalized condition.
Therefore, using $\alpha_{\rm th}(T) \equiv \alpha(T \equiv T_{\rm 21}=T_{\rm 32})$,
the thermalized condition always predicts the minimum CO(2-1) flux of $F_{\rm 21}=(4/9\alpha_{\rm th}(T)) F_{\rm 32}$,
where $\alpha_{\rm th} \approx$1.10, 1.16, and 1.40 for $T$=30, 20, and 10~K, respectively.
Here, we neglected the contribution of the cosmic microwave background radiation since its temperature is low for the CO(3-2)-emitting gas.

The brightest cloud (Cloud 4 in Table \ref{tab:measured}) has the CO(3-2) flux of $F_{32}=258.2\, {\rm mJy}\kmps$.
In the thermalized condition, this translates to the minimum expected CO(2-1) flux of $F_{21}=$126.2, 133.1, and 160.7$\, {\rm mJy}\kmps$
at 30, 20, and 10~K (Eq. \ref{eq:f21}).
This cloud has ${\rm FWHM_v}=6.21\kmps$ in CO(3-2), and if this is also the width in CO(2-1), the minimum predicted CO(2-1) intensity is $I_{21}=$20.3, 21.4, and 25.9$\, {\rm mJy}$.
These are at 1.9, 2.0, and 2.4$\sigma$ significance, respectively.
Thus, it is slightly below the detection limit of the existing CY3NA data.

If the clump is not thermalized, the predicted CO(2-1) flux becomes higher and should have been detected.
For example, if the CO(3-2)-emitting region has $T_{\rm 32}/T_{\rm 21}\sim 0.5$
\citep[i.e., the average in the main disks of nearby galaxies;][]{Leroy:2022ac},
$\alpha(T_{\rm 21}, T_{\rm 32})=\alpha(T, T/2)=$1.51, 1.92, and 4.39 at $T=$30, 20, and 10~K.
Cloud 4 would have $F_{21}=$346.6, 440.7, and 1007.6$\, {\rm mJy}\kmps$ and $I_{21}=$55.8, 71.0, and 162.2$\, {\rm mJy}$.
It should have been detected in CO(2-1) at 5.2, 6.6, and 15.2$\sigma$ in the CY3NA data.
Therefore, the excitation condition of the clump cannot deviate significantly from the thermalized condition,
given the non-detection in CO(2-1).

In addition, if the envelope of the cloud is filled with CO, the CO(2-1) flux would be higher
since the CO(2-1) emission is easily excited in the average density and temperature of typical cloud envelopes.
This should also have made the clump detectable in the previous study.

In summary, the non-detection in CO(2-1) supports the cloud structure presented in Fig. \ref{fig:schematic}b.
This is obviously a simplistic picture, and the details will need some modification in the future.
However, as a first order approximation, Fig. \ref{fig:schematic}b likely represents the cloud structure in the XUV disk.

\section{The size-velocity dispersion relation}

\begin{figure}[h]
    \centering
    \includegraphics[width=0.5\textwidth]{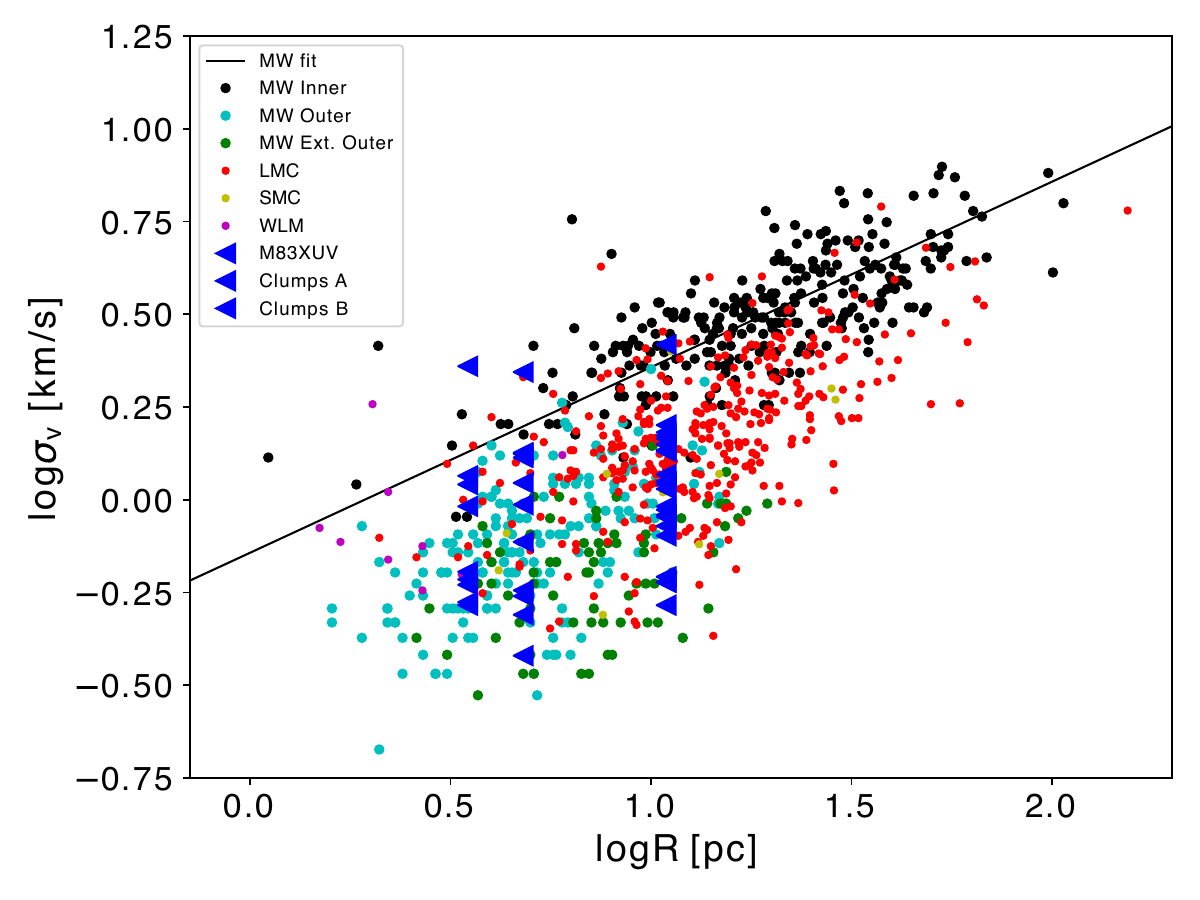}
    \caption{$\sigma_{\rm v}$-$R$ plot.
M83 XUV disk clouds and clumps are in blue (and the blue arrows are the upper limits in radius).\ The clouds have the largest radii, followed by the Level A and then the Level B clumps.
The reference points in the background are for the Milky Way's inner disk \citep{Solomon:1987pr},
outer disk \citep[$r_{\rm gal}\sim$12-14$\kpc$; ][]{Sun:2017vv},
and far outer disk \citep[$\sim$14-22$\kpc$; ][]{Sun:2015wr}, the
LMC and SMC \citep{Wong:2011fk, Bolatto:2008nz},
and WLM \citep{Rubio:2015aa}.
The solid line is a fit to the inner disk clouds \citep{Solomon:1987pr}.}
    \label{fig:R_sigv}
\end{figure}

Figure \ref{fig:R_sigv} plots $\sigma_{\rm v}$ against the upper limits of $R$ for the clouds (arrows at $\log R=1.03$), Level A ($0.68$) and B clumps ($0.54$).
For reference, it also shows clouds in the Milky Way, Large and Small Magellanic Clouds (LMC and SMC), and the dwarf galaxy WLM \citep{Solomon:1987pr, Bolatto:2008nz, Wong:2011fk, Rubio:2015aa}.
The clouds in the XUV disk of M83 appear consistent with those in the dwarfs, while their radii are only upper limits.

There is also a caveat in this comparison, since these reference clouds are observed in the lower-excitation transitions, CO(1-0) or CO(2-1).
However, we note that the three CO transitions ($J$=1-0, 2-1, and 3-2) should be emitted from a similar central region (Fig. \ref{fig:schematic}b) if the clouds and clumps share a similar physical and chemical structure in the low-metallicity environments (i.e., the Milky Way outskirts, LMC, SMC, and other dwarfs, as well as the XUV disk).
In fact, molecular clouds in LMC and SMC appear compact in recent CO(1-0) and CO(2-1) measurements \citep{Harada:2019ab, Tokuda:2021aa, Ohno:2023aa}.
Therefore, the consistency in Fig. \ref{fig:R_sigv} supports our hypothesis (Fig. \ref{fig:schematic}), of course, within the limitation that we have only the upper limits of $R$.

\section{Further discussion}

\citet{Koda:2022aa} presented a hypothesis on the mass structure and CO chemistry in molecular clouds in the XUV disk (Fig. \ref{fig:schematic}).
The Orion A cloud (i.e., a nearby Galactic molecular cloud) exhibits SF activity similar to those around the XUV disk clouds, and it consists of a relatively low-density envelope of $\sim$20-30~pc with a dense CO(3-2)-bright region of $\sim$10~pc in size \citep{Ikeda:1999vt, Nakamura:2019wt}.
The hypothesis suggests that the XUV clouds also have a similar mass structure, but in their envelopes, CO molecules are photo-dissociated and deficient due to the low metallicity condition.
The dissociating UV radiation is in the XUV disk (see \citealt{Koda:2022aa}; there are multiple Orion-like sites of massive SF -- traced by UV and H$\alpha$ sources -- in the $1\kpc^2$ region, i.e., the Cy7 field of view).
This study confirms the compactness of the CO(3-2)-emitting regions ($\lesssim$6-9~pc in diameter) in the XUV disk clouds.
In the Orion A cloud, the dense CO(3-2)-bright region consists of multiple smaller clumps \citep[see ][]{Ikeda:1999vt}, while our resolution is not high enough to resolve such substructures.
This study also shows that the CO(3-2)-emitting regions in the XUV disk clouds must be close to the thermalized condition, since otherwise the expected CO(2-1) flux would be brighter and should have been detected in the past ALMA observations.

\subsection{Mass and radius of the CO(3-2)-emitting regions}
The masses of the clumps (i.e., CO(3-2)-emitting regions) can be crudely estimated \citep{Koda:2022aa}, assuming that
(1) the clumps have CO molecules (as in Fig. \ref{fig:schematic}b),
(2) the Galactic CO-to-H$_2$ conversion factor $X_{\rm CO(1-0)}=3\times 10^{20}\,\rm cm^{-2}\,[K\cdot \kmps]^{-1}$ can be applied to the clumps (i.e., parts of the clouds),\footnote{We assumed this $X_{\rm CO(1-0)}$ for clumps.
If their parental clouds have the structure in Fig. \ref{fig:schematic}b, their envelopes should have a fair amount of mass while no/little CO emission. Thus, $X_{\rm CO(1-0)}$ {for clouds} should be larger than the assumed, which is the often-discussed metallicity dependence of $X_{\rm CO(1-0)}$.
Recently, \citet{Lee:2024aa} also pointed out a possible dependence of $X_{\rm CO(1-0)}$ on cloud population.}
and
(3) the gas is thermalized and hence the CO 3-2/1-0 brightness temperature ratio is unity ($R_{3-2/1-0}=1$).
With these assumptions, the clump mass, including helium and other heavier elements  (i.e., multiplying the total hydrogen mass by 1.37), is
\begin{eqnarray}
    M_{\rm clump}
    &=& 3.6 \times 10^3 \Msun \left(\frac{F_{\rm CO(3-2)} dv}{100\,\rm mJy\cdot \kmps}\right)\left(\frac{D}{4.5\rm \, Mpc}\right)^{2}  \nonumber \\
    &\times & \left( \frac{R_{3-2/1-0}}{1.0} \right)^{-1} \left(\frac{X_{\rm CO(1-0)}}{3.0\times 10^{20}} \right). \label{eq:mclump32}
\end{eqnarray}
The measured CO(3-2) fluxes (Table \ref{tab:measured}) give $M_{\rm clump}=2.1\times 10^2$-$9.3\times 10^3\Msun$.

These $M_{\rm clump}$ can be translated, also crudely, to the radius of clumps (or clump concentrations).
The CO(3-2) excitation requires a density of $n_{\rm H_2}\sim 10^3\,\rm cm^{-3}$, to within an order of magnitude.
By assuming this density as an average density in a spherical CO(3-2)-emitting region,
the clump mass within its radius, including helium and other heavier elements, is
\begin{equation}
    M_{\rm clump} = 3.8\times 10^3\Msun \left( \frac{R}{3\pc} \right)^3 \left( \frac{n_{\rm H_2}}{10^3\,\rm cm^{-3}} \right).
\end{equation}
The range of the clump masses estimated above then corresponds to $R=$1-4~pc.
This is consistent with the new measurements in this study.
The clumps (or their concentrations) are compact and not spatially resolved at the 6-9~pc resolutions,
while there is an indication that they may be starting to be resolved at the 6~pc resolution.

These mass estimations easily have a factor of 2-3 uncertainty.
It is also worth noting that, within such uncertainties, they are roughly consistent with the virial mass $M_{\rm vir}$ \citep[$=9/2 (R \sigma_{\rm v}^2/G)$ for a radial density profile of $\propto 1/r$; see][]{Lee:2024aa}, which is
\begin{equation}
    M_{\rm vir} = 3 \times 10^3 \Msun  \left( \frac{R}{3\pc} \right)  \left( \frac{\sigma_{\rm v}}{1\kmps} \right)^2,
\end{equation}
using rounded values of $R=3\pc$ and $\sigma_{\rm v}=1\kmps$ (see Table \ref{tab:measured}).

\subsection{Cloud evolution and distribution}

The cloud structure in Fig. \ref{fig:schematic} does not take into account the evolution of molecular clouds.
It is possible that the clouds evolve and develop dense clumps
only at a later stage of their lifetime.
In this case, the CO(3-2) observations would selectively trace
the clouds at the evolved stage when they are ready for SF.
In fact, in the inner disk of M83, the populations of star-forming molecular clouds
are mostly localized around the galactic center, bar, and spiral arms.
Molecular gas and clouds are abundant even in the interarm regions,
but appear to be dormant in SF \citep{Koda:2023aa}.
If this is also the case in XUV disks, there could also be (undetected) clouds at an earlier stage of their evolution before the dense clumps develop in their hearts.
In order to understand the SF in XUV disks and in general,
it is important to confirm the presence or absence of such dormant cloud populations
and to characterize their structures and distribution as a function of their environments
(e.g., the amount of the surrounding HI gas, spiral arms, UV radiation field, stellar feedback, etc.).

The large-scale distribution of the clouds may also provide a clue on their formation and evolution \citep[e.g.,][]{Dobbs:2014aa}.
Two additional clouds emerge at the highest sensitivity part of the CY7+9NA data (see Fig. \ref{fig:eximages}b),
while we focused on the clouds and clumps discovered in the CY7NA data.
One is detected at 6.3$\sigma$ between Clouds 9 and 12 at a peak position of (RA, Dec)=(13:37:05.29, -29:59:54.1),
and the other is at 5.3$\sigma$ between Clouds 7 and 8 at (13:37:05.18, -29:59:50.7).
The CY7+9NA data, however, have a nonuniform, asymmetric sensitivity distribution (Fig. \ref{fig:fov}),
and thus, we did not include them in the sample for our analysis.
Nevertheless, they exemplify the fact that more CO(3-2) sources could emerge with longer integration.
\citet{Koda:2022aa} noted that the CO(3-2) sources are distributed along filamentary structures as chains
(like seeds in edamame green soybeans).
The two CY7+9NA clouds appear to support this view.
The distribution may indicate that their formation requires compression due to large-scale gas flows.

\section{Summary}

Massive SF sites have been detected in XUV disks, but their parental gas clouds remained mostly undetected despite searches in CO(1-0) and CO(2-1).
Therefore, the recent discovery of 23 molecular clouds in CO(3-2) in the XUV disk of M83 is startling.
To explain the previous non-detections and the new detections,
\citet{Koda:2022aa} introduced the hypothesis that clouds in XUV disks, like Galactic star-forming clouds, have dense star-forming clumps embedded in a large envelope of bulk molecular gas, where CO molecules are photo-dissociated due to the lack of CO self-shielding and dust extinction in the low-metallicity environment of XUV disks  (Fig. \ref{fig:schematic}a,b).
CO(3-2) can be excited preferentially in the dense clumps but not as much in the less dense envelopes.
This can explain why CO(3-2) is detected when CO(1-0) and CO(2-1) levels are significantly reduced.

This paper confirms the compactness of the dense, CO(3-2)-emitting clumps in the clouds as predicted by this hypothesis.
In the new ALMA data, they are not resolved at a 0.40" (9~pc) resolution (natural weighting),
nor at a 0.28" (6~pc) resolution (uniform weighting),
but there is some indication that they may be starting to be resolved at the 6~pc resolution.
These are similar to the sizes of the dense central clumps (or their concentration) in the Orion A molecular cloud in the solar neighborhood, where a similar amount of massive SF is ongoing.

The CO(3-2) to CO(2-1) and CO(3-2) to CO(1-0) line ratios in the dense clumps must be higher than the average ratio observed in normal galactic disks \citep[e.g., $T_{\rm 32}/T_{\rm 10}\sim 1/3$ and $T_{\rm 32}/T_{\rm 21}\sim 1/2$; ][]{Wilson:2009fk, Vlahakis:2013aa, Leroy:2022ac}.
Otherwise, the previous study \citep{Bicalho:2019aa} would have detected the CO(2-1) emission.
Likewise, the molecular envelopes around the dense clumps within the clouds cannot have much CO, as the CO(1-0) and CO(2-1) emission should be excited easily in the envelopes and, thus, would have been detected.
The CO(3-2)-emitting regions in the XUV clouds must be close to thermalized since, otherwise, the CO(2-1) flux would be brighter and thus would have been detected in past observations.

The cloud structure hypothesized here is, of course, simplistic and will need some adjustments in the future.
For example, it does not include the possible evolution of clouds from a non-star-forming phase to a star-forming phase.
The clouds detected in CO(3-2) may trace only the clouds in the star-forming phase.
For the clouds in the XUV disk discussed in this paper, this model appears to capture the reality within the limitation of the current observations.

\begin{acknowledgements}
JK thanks his colleagues at Paris Observatory for their hospitality, both professionally and personally, during his stay on sabbatical, and colleagues at the Institut d'Astrophysique de Paris for stimulating discussions.
We also thank the anonymous referee.
This paper makes use of the following ALMA data: ADS/JAO.ALMA\#2013.1.00861.S, 2017.1.00065.S, and 2022.1.00359.S.
ALMA is a partnership of ESO (representing its member states), NSF (USA) and NINS (Japan), together with NRC (Canada), MOST and ASIAA (Taiwan), and KASI (Republic of Korea), in cooperation with the Republic of Chile.
The Joint ALMA Observatory is operated by ESO, AUI/NRAO and NAOJ.
Data analysis is in part carried out on the Multi-wavelength Data Analysis System operated by the Astronomy Data Center (ADC) in NAOJ.
J.K. acknowledges support from NSF through grants AST-1812847 and AST-2006600.
M.R. wishes to acknowledge support from ANID(CHILE) through FONDE-CYT grant No1190684.
\end{acknowledgements}

% BibTeX
\bibliographystyle{aa}

\begin{appendix} %First appendix

\section{Parameters and cutout images of clouds and clumps}

\begin{figure*}[h]
    \centering
    \includegraphics[width=1.0\textwidth]{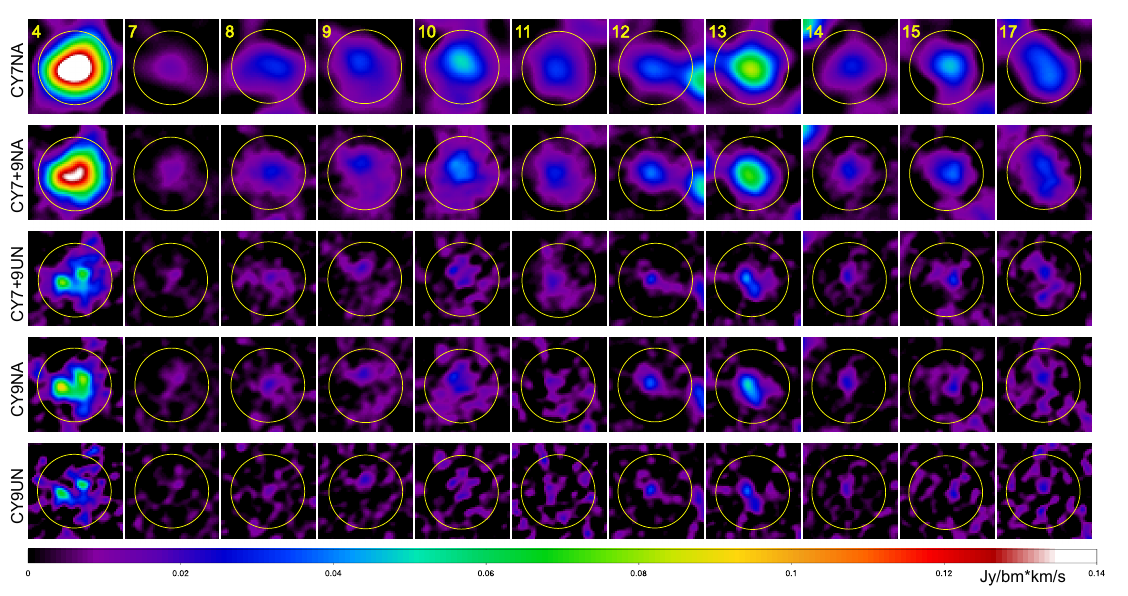}
    \caption{Cutout images of the clouds and clumps in CO(3-2) integrated intensity.
    The velocity range of each cloud is integrated.
    The primary beam correction is applied.
    From left to right: Different clouds, with their ID\# in the top-left corners.
    From top to bottom: CY7NA, CY7+9NA, CY7+9UN, CY9NA, and CY9UN (see Table \ref{tab:data}).
    The yellow circles have a diameter of 2" ($\sim 44$~pc). }
    \label{fig:pstamp0}
\end{figure*}

\begin{figure*}[h]
    \centering
    \includegraphics[width=1.0\textwidth]{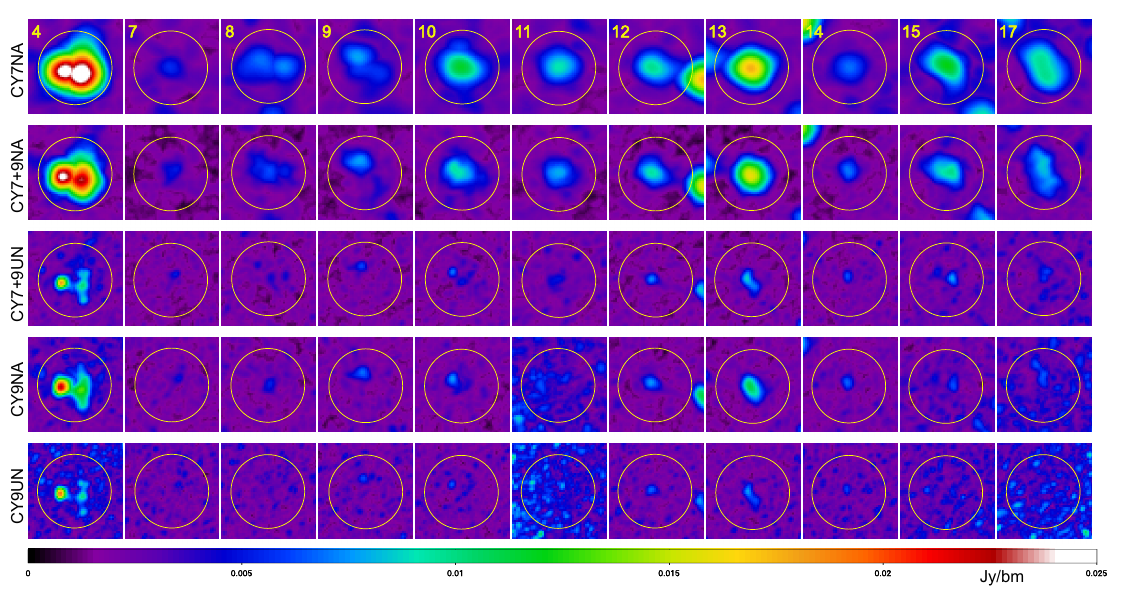}
    \caption{Same as Fig. \ref{fig:pstamp0}, but in peak intensity.}
    \label{fig:pstamp8}
\end{figure*}

Table \ref{tab:measured} lists the measured parameters of the clouds and clumps (see Sect. \ref{sec:parameters} for explanations).
Figures \ref{fig:pstamp0} and \ref{fig:pstamp8} show their cutout images in integrated intensity and peak intensity, respectively.

Table \ref{tab:allclouds} is the same as the "Clouds" section of Table \ref{tab:measured}, but lists all 23 clouds identified in the CY7NA cube.
This table does not include the peak brightness temperature $T_{\rm p}$. The $I_{\rm p}$ and $T_{\rm p}$ are related as $T_{\rm p}=C_{\rm K} I_{\rm p}$ (see Table \ref{tab:data} for $C_{\rm K}$).

\begin{sidewaystable*}
\caption{Measured parameters of clouds and Level A and B clumps in CO(3-2)}             % title of Table
\label{tab:measured}
\centering
\begin{tabular}{cccccccccccccccccc}
\hline\hline                 % inserts double horizontal lines
(1) & (2)  & (3) & (4) & (5) & (6) & (7) & (8) & (9) & \multicolumn{2}{c}{(10)} & (11) & (12) & (13) & (14) & (15) & (16) & (17) \\
\hline
 &  &  &  &  &  & \multicolumn{3}{c}{FWHM} & &  & & &  \\
\cline{7-9}
ID & RA & Dec & $V$ & $N_{\rm pix}$ & $N_{\rm xy}$ & x & y & v  & $T_{\rm p}$ & $I_{\rm p}$  & $\Delta I_{\rm p}$ & $F$ & $\Delta F$ & $R$ & $\sigma_{\rm v}$ & $\Delta \sigma_{\rm v}$ & PB \\
\cline{10-11}
 & \small{J2000} & \small{J2000} & \small{$\kmps$}  &  &  & \small{"} & \small{"} & \small{$\kmps$} & \small{K} & \small{mJy/bm} & \small{mJy/bm} & \small{$\rm mJy \kmps$} & \small{$\rm mJy \kmps$} & \small{"} & \small{$\kmps$} & \small{$\kmps$} & \\
 \hline                        % inserts single horizontal line
\multicolumn{16}{c}{Clouds (CY7NA)}\\
\hline
  %ID &           RA &          DEC &      VEL &    FWHMx &    FWHMy &    FWHMz &     peak &     pkerr &     SumT &   SumTer &  sigmav & sigmaverr &  PBatt \\
  4 &  13:37:04.984 & -29:59:46.15 &  564.4 &  3973 &  1438 &  1.02 &  0.96 &  6.21 & 0.34 & 26.47 &  1.25 &   258.2 &    11.2 & <0.50 &  2.62 &    0.06 &  0.79 \\
   7 &  13:37:05.337 & -29:59:50.60 &  559.8 &   192 &   192 &  0.50 &  0.40 &  1.38 & 0.07 &  5.53 &  1.08 &     5.8 &     2.0 & <0.50 &  0.52 &    0.15 &  0.97 \\
   8 &  13:37:05.054 & -29:59:51.21 &  565.9 &   856 &   684 &  1.07 &  0.58 &  3.23 & 0.09 &  7.26 &  1.17 &    30.5 &     4.7 & <0.50 &  1.35 &    0.08 &  0.88 \\
   9 &  13:37:05.352 & -29:59:52.77 &  561.2 &   904 &   698 &  0.81 &  0.84 &  2.82 & 0.10 &  7.80 &  1.08 &    30.2 &     4.4 & <0.50 &  1.17 &    0.08 &  0.97 \\
  10 &  13:37:05.040 & -29:59:53.04 &  567.5 &  1266 &   935 &  0.80 &  1.04 &  3.36 & 0.15 & 11.64 &  1.17 &    49.6 &     5.7 & <0.50 &  1.41 &    0.10 &  0.87 \\
  11 &  13:37:05.917 & -29:59:55.36 &  564.4 &   752 &   572 &  0.71 &  0.71 &  2.21 & 0.12 &  9.13 &  1.06 &    27.1 &     3.9 & <0.50 &  0.91 &    0.13 &  0.99 \\
  12 &  13:37:05.186 & -29:59:55.41 &  565.5 &   700 &   494 &  0.74 &  0.58 &  2.60 & 0.13 &  9.74 &  1.11 &    27.4 &     4.0 & <0.50 &  1.07 &    0.11 &  0.93 \\
  13 &  13:37:05.084 & -29:59:55.69 &  567.6 &  1607 &   748 &  0.82 &  0.69 &  3.64 & 0.23 & 17.57 &  1.16 &    78.6 &     6.4 & <0.50 &  1.52 &    0.08 &  0.88 \\
  14 &  13:37:04.978 & -29:59:56.98 &  567.8 &   388 &   313 &  0.59 &  0.49 &  2.34 & 0.09 &  6.81 &  1.22 &    13.3 &     3.3 & <0.50 &  0.96 &    0.17 &  0.85 \\
  15 &  13:37:04.820 & -29:59:59.49 &  570.4 &   763 &   607 &  0.75 &  0.78 &  2.32 & 0.16 & 12.09 &  1.34 &    35.8 &     5.2 & <0.50 &  0.94 &    0.11 &  0.75 \\
  17 &  13:37:04.738 & -30:00:01.06 &  569.6 &   754 &   626 &  0.74 &  0.83 &  2.06 & 0.13 &  9.76 &  1.42 &    35.6 &     5.5 & <0.50 &  0.85 &    0.10 &  0.70 \\
  \hline
\multicolumn{16}{c}{Level A Clumps (CY9NA)}\\
\hline
 4-1 &  13:37:05.010 & -29:59:46.20 &  563.0 &   230 &   154 &  0.33 &  0.39 &  3.21 & 1.46 & 22.21 &  1.47 &    79.3 &     7.1 & <0.22 &  1.34 &    0.09 &  0.63 \\
 4-2 &  13:37:04.971 & -29:59:46.22 &  565.8 &   452 &   250 &  0.32 &  0.66 &  5.23 & 0.88 & 13.40 &  1.49 &   116.7 &    10.3 & <0.22 &  2.21 &    0.07 &  0.61 \\
 7-1 &  13:37:05.325 & -29:59:50.43 &  559.8 &    31 &    31 &  0.20 &  0.17 &  1.46 & 0.32 &  4.92 &  1.06 &     4.4 &     1.8 & <0.22 &  0.57 &    0.16 &  0.92 \\
 8-1 &  13:37:05.049 & -29:59:51.22 &  565.0 &    64 &    61 &  0.23 &  0.28 &  1.90 & 0.34 &  5.21 &  1.04 &     9.3 &     2.5 & <0.22 &  0.77 &    0.21 &  0.94 \\
 9-1 &  13:37:05.363 & -29:59:52.46 &  559.8 &   103 &   103 &  0.40 &  0.27 &  1.43 & 0.45 &  6.79 &  1.02 &    17.0 &     3.2 & <0.22 &  0.55 &    0.08 &  0.95 \\
10-1 &  13:37:05.055 & -29:59:52.89 &  567.2 &    94 &    91 &  0.29 &  0.32 &  2.37 & 0.50 &  7.56 &  1.00 &    15.9 &     2.9 & <0.22 &  0.97 &    0.17 &  0.97 \\
10-2 &  13:37:05.032 & -29:59:53.00 &  565.7 &    61 &    49 &  0.22 &  0.27 &  3.12 & 0.27 &  4.17 &  1.01 &     7.7 &     2.4 & <0.22 &  1.30 &    0.20 &  0.97 \\
12-1 &  13:37:05.195 & -29:59:55.34 &  565.7 &   169 &   122 &  0.33 &  0.32 &  2.69 & 0.50 &  7.59 &  1.00 &    28.0 &     3.9 & <0.22 &  1.11 &    0.10 &  0.98 \\
13-1 &  13:37:05.086 & -29:59:55.76 &  567.6 &   391 &   246 &  0.45 &  0.49 &  3.17 & 0.73 & 11.15 &  1.01 &    74.5 &     6.1 & <0.22 &  1.33 &    0.10 &  0.95 \\
14-1 &  13:37:04.982 & -29:59:56.87 &  567.4 &    57 &    57 &  0.21 &  0.27 &  1.30 & 0.44 &  6.64 &  1.09 &     9.7 &     2.5 & <0.22 &  0.49 &    0.11 &  0.90 \\
15-1 &  13:37:04.806 & -29:59:59.42 &  570.0 &    40 &    40 &  0.16 &  0.26 &  1.08 & 0.42 &  6.38 &  1.42 &     7.4 &     2.7 & <0.22 &  0.38 &    0.23 &  0.68 \\
\hline
\multicolumn{16}{c}{Level B Clumps (CY9UN)}\\
\hline
 4-1-1 &  13:37:05.010 & -29:59:46.22 &  562.7 &    77 &    61 &  0.26 &  0.22 &  2.32 & 2.49 & 19.04 &  1.92 &    55.0 &     7.4 & <0.16 &  0.96 &    0.09 &  0.65 \\
 4-2-1 &  13:37:04.970 & -29:59:46.30 &  564.9 &    39 &    39 &  0.18 &  0.22 &  1.62 & 1.29 &  9.83 &  1.94 &    21.3 &     5.3 & <0.16 &  0.64 &    0.04 &  0.65 \\
 4-2-2 &  13:37:04.968 & -29:59:46.69 &  564.9 &    22 &    22 &  0.14 &  0.16 &  1.55 & 1.12 &  8.54 &  1.87 &    10.9 &     3.8 & <0.16 &  0.61 &    0.06 &  0.68 \\
 4-2-3 &  13:37:04.969 & -29:59:45.99 &  567.4 &    34 &    25 &  0.16 &  0.14 &  5.42 & 1.15 &  8.76 &  2.02 &    17.3 &     5.1 & <0.16 &  2.29 &    0.14 &  0.63 \\
 9-1-1 &  13:37:05.357 & -29:59:52.41 &  559.8 &    23 &    23 &  0.17 &  0.14 &  1.49 & 0.81 &  6.16 &  1.33 &     8.2 &     2.8 & <0.16 &  0.59 &    0.11 &  0.96 \\
10-1-1 &  13:37:05.059 & -29:59:52.83 &  567.4 &    17 &    17 &  0.13 &  0.13 &  1.34 & 0.85 &  6.49 &  1.31 &     6.3 &     2.3 & <0.16 &  0.52 &    0.14 &  0.98 \\
12-1-1 &  13:37:05.194 & -29:59:55.37 &  565.9 &    51 &    31 &  0.18 &  0.17 &  2.80 & 0.88 &  6.70 &  1.30 &    19.0 &     4.0 & <0.16 &  1.16 &    0.15 &  0.98 \\
13-1-1 &  13:37:05.089 & -29:59:55.76 &  567.7 &    97 &    80 &  0.27 &  0.41 &  2.63 & 1.00 &  7.63 &  1.32 &    36.9 &     5.7 & <0.16 &  1.10 &    0.16 &  0.96 \\
14-1-1 &  13:37:04.983 & -29:59:56.90 &  567.4 &    14 &    14 &  0.09 &  0.15 &  1.42 & 0.78 &  5.99 &  1.42 &     5.2 &     2.3 & <0.16 &  0.53 &    0.14 &  0.90 \\
\hline                                   %inserts single line
 \end{tabular}
 \tablefoot{
 (1) Cloud ID.
 (2)(3)(4) Coordinate and recession velocity $V$.
 (5)(6) Numbers of pixels identified within cloud in the RA-Dec-Vel space $N_{\rm pix}$ and in the RA-Dec projection $N_{\rm xy}$.
 (7)(8)(9) FWHM widths in RA (=x), Dec (=y), and velocity (=v) directions, calculated from intensity-weighted dispersions. The FWHMv is calculated with the $0.85\kmps$-channel cube, while the rest is with the $2.54\kmps$-channel cube.
 (10)(11) Peak intensity $I_{\rm p}$ and uncertainty $\Delta I_{\rm p}$ after primary beam correction.
 $I_{\rm p}$ is also given in the form of brightness temperature $T_{\rm p}$.
 (12)(13) Total flux $F$ and uncertainty $\Delta F$ after primary beam correction, calculated by summing up the fluxes of all the pixels identified in cloud or clump.
 (14) Radius after beam dilution correction. Unlike $\sigma_{\rm r}$, $R$ is supposed to trace the whole radial extent of object.
 (15)(16) Velocity dispersion and uncertainty after channel dilution correction.
 (17) Primary beam attenuation.
}                                                                                  \end{sidewaystable*}

\clearpage

\clearpage
\begin{sidewaystable*}
\caption{Cloud parameter measurements in CO(3-2) with the CY7NA cube.}             % title of Table
\label{tab:allclouds}
\centering
\begin{tabular}{ccccccccccccccccc}
\hline\hline                 % inserts double horizontal lines
(1) & (2)  & (3) & (4) & (5) & (6) & (7) & (8) & (9) & (10) & (11) & (12) & (13) & (14) & (15) & (16) & (17) \\
\hline
 &  &  &  &  &  & \multicolumn{3}{c}{FWHM} & &  & & &  \\
\cline{7-9}
ID & RA & Dec & $V$ & $N_{\rm pix}$ & $N_{\rm xy}$ & x & y & v  & $I_{\rm p}$ & $\Delta I_{\rm p}$ & $F$ & $\Delta F$ & $R$ & $\sigma_{\rm v}$ & $\Delta \sigma_{\rm v}$ & PB \\
 & \small{J2000} & \small{J2000} & \small{$\kmps$}  &  &  & \small{"} & \small{"} & \small{$\kmps$} & \small{mJy/bm} & \small{mJy/bm} & \small{$\rm mJy \kmps$} & \small{$\rm mJy \kmps$} & \small{"} & \small{$\kmps$} & \small{$\kmps$} & \\
 \hline                        % inserts single horizontal line
   1 &  13:37:05.044 & -29:59:33.96 &  558.6 &   679 &   456 &  0.85 &  0.48 &  3.24 & 18.37 &  3.50 &    67.5 &    13.7 & <0.50 &  1.36 &    0.03 &  0.27 \\
   2 &  13:37:05.035 & -29:59:36.15 &  555.3 &   364 &   281 &  0.55 &  0.49 &  3.50 & 13.10 &  2.51 &    24.0 &     6.8 & <0.50 &  1.46 &    0.10 &  0.40 \\
   3 &  13:37:04.814 & -29:59:40.69 &  561.6 &  1471 &   945 &  0.94 &  0.86 &  3.55 & 17.71 &  1.90 &   100.2 &    10.8 & <0.50 &  1.49 &    0.06 &  0.50 \\
   4 &  13:37:04.984 & -29:59:46.15 &  564.4 &  3973 &  1438 &  1.02 &  0.96 &  6.21 & 26.47 &  1.25 &   258.2 &    11.2 & <0.50 &  2.62 &    0.06 &  0.79 \\
   5 &  13:37:06.467 & -29:59:49.13 &  562.2 &   618 &   574 &  0.82 &  0.83 &  1.96 &  8.22 &  1.12 &    22.7 &     3.8 & <0.50 &  0.80 &    0.12 &  0.92 \\
   6 &  13:37:06.179 & -29:59:50.41 &  552.2 &   508 &   508 &  1.00 &  0.54 &  1.52 &  6.37 &  1.07 &    16.5 &     3.2 & <0.50 &  0.60 &    0.04 &  0.98 \\
   7 &  13:37:05.337 & -29:59:50.60 &  559.8 &   192 &   192 &  0.50 &  0.40 &  1.38 &  5.53 &  1.08 &     5.8 &     2.0 & <0.50 &  0.52 &    0.15 &  0.97 \\
   8 &  13:37:05.054 & -29:59:51.21 &  565.9 &   856 &   684 &  1.07 &  0.58 &  3.23 &  7.26 &  1.17 &    30.5 &     4.7 & <0.50 &  1.35 &    0.08 &  0.88 \\
   9 &  13:37:05.352 & -29:59:52.77 &  561.2 &   904 &   698 &  0.81 &  0.84 &  2.82 &  7.80 &  1.08 &    30.2 &     4.4 & <0.50 &  1.17 &    0.08 &  0.97 \\
  10 &  13:37:05.040 & -29:59:53.04 &  567.5 &  1266 &   935 &  0.80 &  1.04 &  3.36 & 11.64 &  1.17 &    49.6 &     5.7 & <0.50 &  1.41 &    0.10 &  0.87 \\
  11 &  13:37:05.917 & -29:59:55.36 &  564.4 &   752 &   572 &  0.71 &  0.71 &  2.21 &  9.13 &  1.06 &    27.1 &     3.9 & <0.50 &  0.91 &    0.13 &  0.99 \\
  12 &  13:37:05.186 & -29:59:55.41 &  565.5 &   700 &   494 &  0.74 &  0.58 &  2.60 &  9.74 &  1.11 &    27.4 &     4.0 & <0.50 &  1.07 &    0.11 &  0.93 \\
  13 &  13:37:05.084 & -29:59:55.69 &  567.6 &  1607 &   748 &  0.82 &  0.69 &  3.64 & 17.57 &  1.16 &    78.6 &     6.4 & <0.50 &  1.52 &    0.08 &  0.88 \\
  14 &  13:37:04.978 & -29:59:56.98 &  567.8 &   388 &   313 &  0.59 &  0.49 &  2.34 &  6.81 &  1.22 &    13.3 &     3.3 & <0.50 &  0.96 &    0.17 &  0.85 \\
  15 &  13:37:04.820 & -29:59:59.49 &  570.4 &   763 &   607 &  0.75 &  0.78 &  2.32 & 12.09 &  1.34 &    35.8 &     5.2 & <0.50 &  0.94 &    0.11 &  0.75 \\
  16 &  13:37:06.684 & -29:59:59.73 &  564.9 &   513 &   387 &  0.54 &  0.69 &  3.79 &  6.18 &  1.22 &    16.6 &     3.8 & <0.50 &  1.59 &    0.17 &  0.85 \\
  17 &  13:37:04.738 & -30:00:01.06 &  569.6 &   754 &   626 &  0.74 &  0.83 &  2.06 &  9.76 &  1.42 &    35.6 &     5.5 & <0.50 &  0.85 &    0.10 &  0.70 \\
  18 &  13:37:06.407 & -30:00:01.43 &  564.0 &   465 &   329 &  0.54 &  0.57 &  2.71 &  7.40 &  1.09 &    15.0 &     3.2 & <0.50 &  1.12 &    0.14 &  0.96 \\
  19 &  13:37:06.006 & -30:00:02.05 &  562.3 &   410 &   410 &  0.62 &  0.74 &  1.58 &  6.00 &  1.06 &    13.1 &     2.9 & <0.50 &  0.62 &    0.04 &  1.00 \\
  20 &  13:37:04.848 & -30:00:03.57 &  569.0 &   681 &   544 &  0.88 &  0.62 &  2.68 &  7.43 &  1.30 &    24.3 &     4.6 & <0.50 &  1.12 &    0.10 &  0.79 \\
  21 &  13:37:05.567 & -30:00:06.58 &  567.4 &   230 &   230 &  0.64 &  0.36 &  1.54 &  6.07 &  1.07 &     7.3 &     2.2 & <0.50 &  0.60 &    0.06 &  0.99 \\
  22 &  13:37:05.770 & -30:00:11.22 &  570.5 &   547 &   533 &  1.03 &  0.58 &  2.65 &  8.34 &  1.16 &    19.7 &     3.7 & <0.50 &  1.10 &    0.12 &  0.89 \\
  23 &  13:37:05.209 & -30:00:15.15 &  567.1 &   230 &   198 &  0.49 &  0.41 &  2.20 &  8.05 &  1.56 &     9.8 &     3.3 & <0.50 &  0.90 &    0.18 &  0.65 \\
\hline                                   %inserts single line
 \end{tabular}
 \tablefoot{
 (1) Cloud ID.
 (2)(3)(4) Coordinate and recession velocity $V$.
 (5)(6) Numbers of pixels identified within cloud in the RA-Dec-Vel space $N_{\rm pix}$ and in the RA-Dec projection $N_{\rm xy}$.
 (7)(8)(9) FWHM widths in RA (=x), Dec (=y), and velocity (=v) directions, calculated from intensity-weighted dispersions. The FWHMv is calculated with the $0.85\kmps$-channel cube, while the rest is with the $2.54\kmps$-channel cube.
 (10)(11) Peak intensity $I_{\rm p}$ and uncertainty $\Delta I_{\rm p}$ after primary beam correction.
 (12)(13) Total flux $F$ and uncertainty $\Delta F$ after primary beam correction, calculated by summing up the fluxes of all the pixels identified in cloud.
 (14) Radius after beam dilution correction. Unlike $\sigma_{\rm r}$, $R$ is supposed to trace the whole radial extent of object.
 (15)(16) Velocity dispersion and uncertainty after channel dilution correction.
 (17) Primary beam attenuation.
}                                                                                                                                              \end{sidewaystable*}

\clearpage
\section{The stacked cutout images} \label{sec:stack}

A stacking analysis does not provide a significant CO(2-1) detection in this case.
The clouds have a large range of flux (Table \ref{tab:measured}),
and only the brightest ones contribute to the stacked flux, while all the cloud images contribute to the stacked noise.

\begin{figure}[h]
    \centering
    \includegraphics[width=0.5\textwidth]{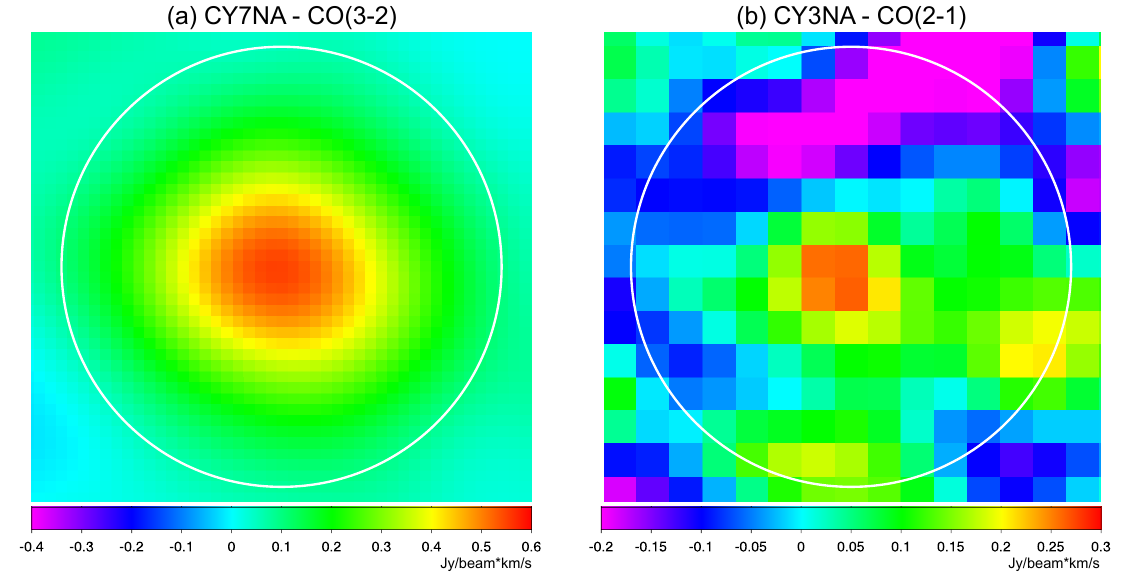}
    \caption{Stacked (summed) images of the 11 clouds in the Cy9 FoV (see Fig. \ref{fig:fov} and Table \ref{tab:measured}).
    (a) CO(3-2) image with CY7NA (with an RMS noise of 16$\, {\rm mJy/beam}\kmps$).
    (b) CO(2-1) image with CY3NA (140$\, {\rm mJy/beam}\kmps$).
    The white circles have a diameter of 2".
    Note that the beam sizes are different in the two panels (see Table \ref{tab:data}).
    Assuming that the emission is not resolved, the fluxes in these images indicate the total fluxes of the (stacked) object.
    }
    \label{fig:pstampstacks}
\end{figure}

Figure \ref{fig:pstampstacks} shows the stacked CO(3-2) and CO(2-1) images, that is, the sum of the integrated intensity cutout images of the 11 clouds (see Figs. \ref{fig:fov} and \ref{fig:pstamp0} and Table \ref{tab:measured}).
The same velocity ranges are integrated for CO(3-2) and CO(2-1).
The beam sizes are different between the two panels (Table \ref{tab:data}).
If we assume that the (stacked) object is not resolved with either beam, the total flux of the object is the values displayed in these images.

Panel (a) shows the peak CO(3-2) flux of $F_{32}=550\, {\rm mJy} \kmps$.
The noise of this image is $16\, {\rm mJy} \kmps$.
The corresponding CO(2-1) flux, predicted under the thermalized condition, is $F_{21}=(4/9\alpha_{\rm th})F_{32}=269\, {\rm mJy} \kmps$ ($\alpha_{\rm th}\approx 1.1$ at $T$=30~K; see Sect. \ref{sec:excitation} and Eq. \ref{eq:f21}).

Panel (b) shows a tempting emission peak, but there is also a similarly deep negative just north of it.
Indeed, the peak flux is $F_{21}=263\, {\rm mJy}\kmps$ in CO(2-1), which is apparently consistent with the prediction under the thermalized condition.
However, the noise in this image is $140\, {\rm mJy} \kmps$.
Thus, it is only at a $1.9\sigma$ significance.
The consistency is supportive of the model (Fig. \ref{fig:schematic}) but is not conclusive with this CO(2-1) data.

\end{appendix}

\end{document}